\documentstyle[epsf,preprint]{jpsj}
\def\al{\alpha}
\def\be{\beta}
\def\gamma{\ga}
\def\de{\delta}
\def\d{{\rm d}}
\def\lan{\left\langle}
\def\ran{\right\rangle}

\def\e{{\rm e}}
\def\virg{,}
\def\point{.}
\def\vf{v_{\rm F}}
\def\kf{k_{\rm F}}

\def\ggs{\buildrel\textstyle > \over {\hbox{\raise0.2ex\hbox{$\sim$}}}}
\def\lls{\buildrel\textstyle < \over {\hbox{\raise0.2ex\hbox{$\sim$}}}}
\def\gsim{\,\lower0.75ex\hbox{$\ggs$}\,}
\def\lsim{\,\lower0.75ex\hbox{$\lls$}\,}
\def\N{\hat{N}}

\def\im{{\rm i}}
\def\ie{{\it i.e.}, }
\def\delx{\partial_x}
\def\deltau{\partial_\tau}
\def\en{\epsilon_n}
\def\et{{\it{et al.} }}
\def\jo #1#2#3#4{#1 {\bf #2} (#3) #4}   

\def\PRB{Phys.\ Rev.\ B}
\def\PRL{Phys.\ Rev.\ Lett.}

\def\SSC{Solid State Commun.}

\def\JPSJ{J.\ Phys.\ Soc.\ Jpn.}

\def\PTP{Prog.\ Theor.\ Phys.}

\def\ADV{Adv.\ Phys.}

\def\EPL{Europhys.\ Lett.}

\def\EPJ{Eur.\ Phys.\ J.\ B}
\def\runtitle{
 Effects of Next-Nearest-Neighbor Repulsion on 
 One-Dimensional Quarter-Filled Electron Systems 
}
\def\runauthor{Hideo Yoshioka, Masahisa Tsuchiizu,Yoshikazu Suzumura}

\title
{
 Effects of Next-Nearest-Neighbor Repulsion on 
 One-Dimensional Quarter-Filled Electron Systems
}
\author{
H. Yoshioka$^a$\hspace{-1.5mm}
\thanks{
E-mail : yoshioka@phys.nara-wu.ac.jp
}
M. Tsuchiizu$^b$ and Y. Suzumura$^{b,c}$
}

\inst{
$^a$Department of Physics, Nara Women's University, Nara 630-8506 \\
$^b$Department of Physics, Nagoya University, Nagoya 464-8602 \\
$^c$CREST, Japan Science and Technology Corporation (JST)
}

\recdate{September 12, 2000}

\abst{
We examine effects of the next-nearest-neighbor repulsion
on electronic states of a 
  one-dimensional interacting electron system 
which consists of 
quarter-filled band and interactions of on-site and 
nearest-neighbor repulsion.
We derive the effective Hamiltonian for the electrons around wave number 
 $\pm \kf$
($\kf$: Fermi wave number)
and apply the renormalization group method to the  bosonized 
Hamiltonian. 
It is shown that the next-nearest-neighbor repulsion 
makes $4\kf$-charge ordering unstable and suppresses the spin fluctuation.
Further the excitation gaps and spin susceptibility are also evaluated.
}

\kword{organic compounds, charge density waves, 
spin density waves, quarter-filling, long-range Coulomb interaction, 
spin gap} 

\begin{document}
\sloppy
\maketitle

\section{Introduction}

Quasi-one-dimensional organic conductors, 
many of which are the 2:1 charge transfer salts, 
exhibit a variety of physical properties
depending on temperature, pressure and so on\cite{review}. 
A one-dimensional (1-D) interacting electron system at quarter-filling
is a basic model for understanding such exotic properties.  
The electronic states of these materials have been 
often theoretically investigated by use of    
the model with only the on-site interaction $U$ ($> 0$). 
Such a model with the on-site repulsion 
has been used  to explain 
spin density wave (SDW) states observed experimentally.  

Recent experiments on these materials indicate a coexistence of 
SDW with charge density wave (CDW)  
\cite{Pouget-Ravy,Hiraki-Kanoda,Kagoshima,Nogami}.
In (TMTSF)$_2$PF$_6$, 2$\kf$-SDW state coexists with 
2$\kf$- and  4$\kf$-CDW 
below the SDW transition temperature\cite{Pouget-Ravy}, while 
the two kinds of CDW's disappear below 3-4 K\cite{Kagoshima}. 
The material (DI-DCNQI)$_2$Ag shows the magnetic order of 
2$\kf$-SDW  below 5.5 K\cite{Hiraki-Kanoda-II}. 
However, the analysis 
of $^{13} $C-NMR spectra\cite{Hiraki-Kanoda} 
suggests the 4$\kf$-CDW ordering 
below 220 K, which is much higher than the transition temperature 
of the magnetic order.  
Such a 4$\kf$-CDW order has been also observed in 
X-ray study.\cite{Nogami} 
It is known that 
the  coexistence of SDW and CDW observed 
experimentally cannot be explained 
by a model with only the on-site repulsion.  
Therefore, 
it is necessary to treat models with 
the long-range components of 
Coulomb interaction, which may take crucial roles for 
these coexistent states. 

Effects of  several long-range interactions
have been theoretically studied especially for 
a 1-D quarter-filled interacting electron system, 
and the rich phase diagram has been found. 
Numerical diagonalization of the model with 
   both $U$ and  the nearest-neighbor interaction, $V_1$,
\cite{Mila-Zotos,Penc-Mila-I,Sano-Ono,Nakamura-Kitazawa-Nomura}       
 shows that 
 the insulating phase appears for large strengths of both $U$ and
$V_1$,   
 and that the superconducting state 
becomes the most dominant fluctuation
   for large $V_1$ and small $U$.
The mean-field theories have predicted the some coexistent states of SDW 
and CDW.\cite{Seo-Fukuyama,Suzumura,Kobayashi-Ogata-Yonemitsu,Tomio-Suzumura}  
For the system with $U$ and $V_1$, a transition occurs 
 from a pure $2\kf$-SDW state to 
   a  coexistent state of $2\kf$-SDW and 
     $4\kf$-CDW with increasing $V_1$.
\cite{Seo-Fukuyama}
It is maintained that such a coexistent state gives rise to 
   the  charge ordering in (DI-DCNQI)$_2$Ag\cite{Hiraki-Kanoda}.
 The transition has been examined 
  by evaluating the commensurability energy
     corresponding to the $8\kf$-Umklapp scattering.\cite{Suzumura} 
Kobayashi \et found that the 
next-nearest-neighbor repulsion, $V_2$,  results in  
    a coexistence of $2\kf$-SDW and purely electronic $2\kf$-CDW.
\cite{Kobayashi-Ogata-Yonemitsu}
The problem has been reexamined by Tomio and 
one of the present authors\cite{Tomio-Suzumura} 
and it has been found that 
the coexistent state of $2\kf$-SDW and $2\kf$-CDW is followed by 
$4\kf$-SDW but without $4\kf$-CDW. 
The next-nearest-neighbor repulsion
is the most promising candidate 
for the origin of the coexistence   
 observed in (TMTSF)$_2$PF$_6$.
\cite{Pouget-Ravy,Kagoshima} 

In the previous works, 
we have analytically derived the $8\kf$-Umklapp scattering 
for the electron system at quarter-filling with $U$ and $V_1$, 
and investigated the    
electronic properties
    by applying renormalization group (RG)
      method to the bosonized model\cite{Tsuchiizu,Yoshioka}.  
The phase diagram has been determined  
on the plane of $U$ and $V_1$, 
which is qualitatively the same 
as that derived by the numerical approach. 
We found that   
the insulating state  for both large $U$ and $V_1$ 
 exhibits the order of $4\kf$-CDW with
 the most dominant fluctuation given by $2\kf$-SDW. 
Such a state is 
similar to the coexistent state predicted 
by the mean-field theory and seems to  
correspond to the experimental observation in 
(DI-DCNQI)$_2$Ag that 
the characteristic temperature of the charge ordering 
is much larger than that of 2$\kf$-SDW. 

In the present paper, 
we extend the previous method\cite{Tsuchiizu,Yoshioka} 
to the system with $U$, $V_1$ 
and $V_2$, and 
clarify the role of $V_2$ for the electronic states 
and the excitations.\cite{Yoshioka-II}  
We find that $V_2$ makes 4$\kf$-CDW state unstable and 
suppresses the spin fluctuation. 
The phase diagram is determined and compared with that by the mean-field
theory. 
Electronic states are further examined by evaluating 
the excitation gaps of the charge and spin fluctuation, and 
spin susceptibility. 

The plan of the paper is as follows. 
In \S2, the effective Hamiltonian is derived and 
expressed in terms of bosonization with the phase variables. 
The phase diagram and excitations are determined by utilizing  
RG method in \S3.
Based on the RG equations, the spin susceptibility is calculated 
in \S4.
Section 5 is devoted to summary.   
 
\section{Model and Formulation} 
\subsection{Effective Hamiltonian}

We consider 
 a 1-D electron system at quarter-filling.  
The Hamiltonian is given by 
 ${\cal H} = {\cal H}_0 + {\cal H}_{\rm int}$,  
where 
\begin{eqnarray}
{\cal H}_0 
&=& - t \sum_{j \sigma} 
   \left( a_{j, \sigma}^\dagger  a_{j+1, \sigma}  + {\rm h.c.}\right) 
   -\mu \sum_{j, \sigma} n_{j, \sigma}  \nonumber \\
&=& \sum_{K \sigma} (\epsilon_K - \mu ) a^\dagger_{K, \sigma} a_{K, \sigma}  \virg 
\label{eqn:H0}\\
{\cal H}_{\rm int} &=& \sum_{j \sigma \sigma'}
\left\{
\frac{U}{2} \delta_{\sigma' -\sigma}
n_{j, \sigma} n_{j, \sigma'} 
+ V_1  n_{j, \sigma} n_{j+1, \sigma'} + V_2  n_{j, \sigma} n_{j+2, \sigma'}  
\right\}  
\nonumber \\
&=& \frac{1}{N_L} \sum_{\sigma \sigma'} \sum_{K_1 \sim K_4}
\left\{
\frac{U}{2} \delta_{\sigma, -\sigma'} 
+ V_1 {\rm e}^{- \im (K_2 - K_3)a} 
+ V_2 {\rm e}^{- \im 2 (K_2 - K_3)a}  
\right\} \nonumber \\
&\times& \delta_{K_1 + K_2 - K_3 - K_4, G}
a^\dagger_{K_1, \sigma} a^\dagger_{K_2, \sigma'} a_{K_3, \sigma'} a_{K_4, \sigma} \point 
\label{eqn:Hint}
\end{eqnarray}
Here $t$ and $\mu$ denote 
  a transfer energy between the nearest-neighbor sites and   
a  chemical potential, respectively,   
 $\epsilon_K = -2t \cos Ka$ with a lattice constant $a$,  and  
$- \pi/a < K \leq \pi/a$.
The quantity 
$a_{j, \sigma}^\dagger ( = 1/\sqrt{N_L} \sum_K \e^{-\im K a j} 
a_{K, \sigma}^\dagger) $ is a creation
operator of the electron at the $j$-th site with spin
$\sigma(=\pm)$,   
$n_{j, \sigma} = a_{j, \sigma}^\dagger a_{j, \sigma}$,  
  $G = 0$ and $\pm 2 \pi/a$, and   
 $N_L$ is the number of the lattice.

\begin{figure}
\centerline{\epsfxsize=9.0cm\epsfbox{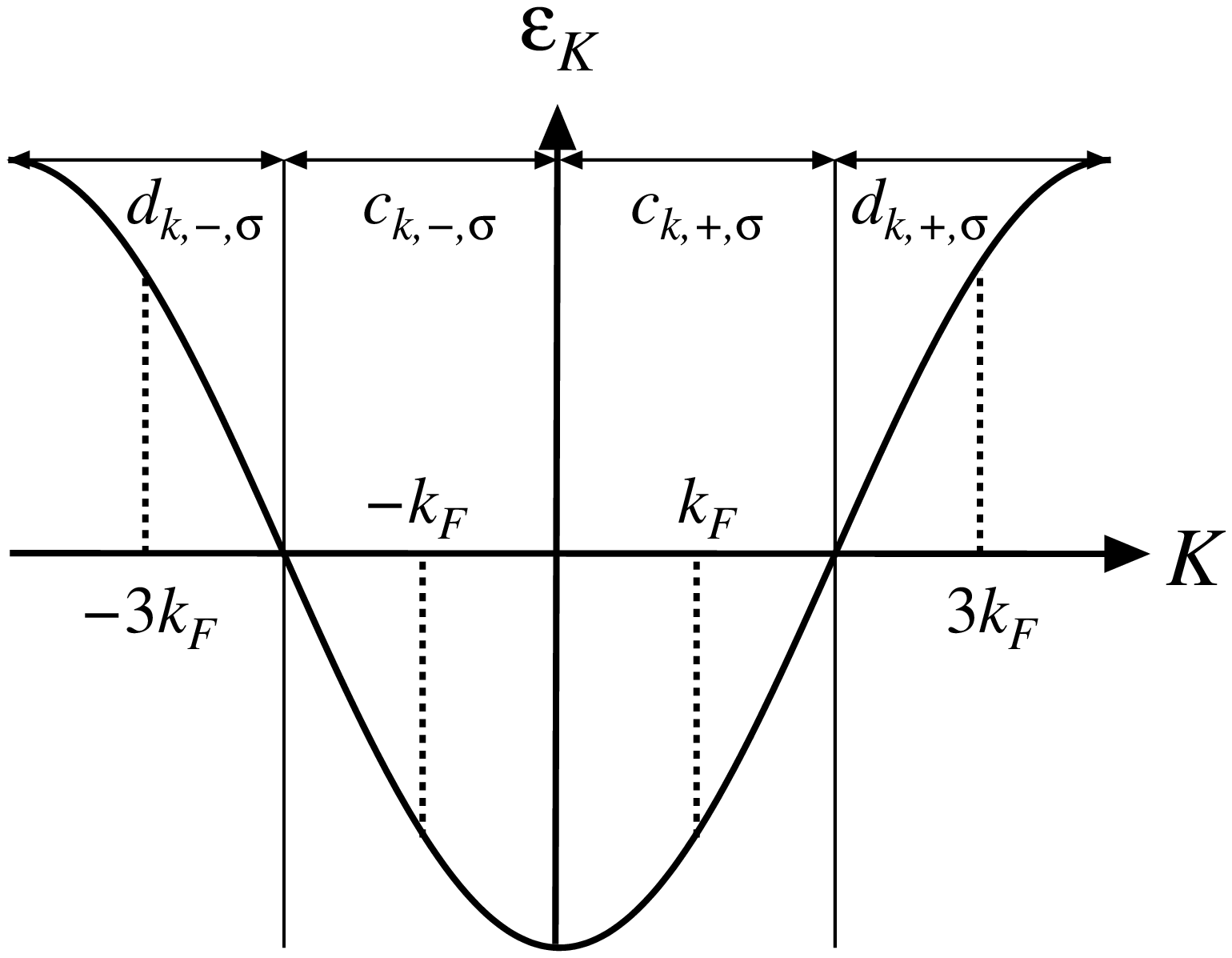}}
\caption{ 
The energy dispersion, $\epsilon_K = -2t \cos Ka$, 
 in the present system. 
}
\label{fig:band}
\end{figure}

In order to obtain the effective Hamiltonian for 
the states near $\pm \kf$ ($\kf = \pi/(4a)$),   
the one-particle states are divided as shown in Fig.~\ref{fig:band}, 
\ie 
$d_{k, -, \sigma} = a_{K, \sigma}$ for $- \pi /a < K \leq - \pi / (2a)$, 
$c_{k, -, \sigma} = a_{K, \sigma}$ for $- \pi /(2a) < K \leq 0$,
$c_{k, +, \sigma} = a_{K, \sigma}$ for $0 < K \leq \pi/(2a)$, and
$d_{k, +, \sigma} = a_{K, \sigma}$ for $\pi /(2a) < K \leq \pi / a$,
where $k$ is deviation of the wave number from $\pm \kf$ for $c_{k, \pm,
\sigma}$, and $\pm 3\kf$ for $d_{k, \pm, \sigma}$.
Hereafter the one-particle states expressed by $d_{k,\pm,\sigma}$
($c_{k,\pm,\sigma}$) are called as the upper (lower) band. 
In terms of $c_{k, p, \sigma}$ and $d_{k, p, \sigma}$ ($p=\pm$), 
${\cal H}$ is written as 
${\cal H} = {\cal H}_0 + \sum_{i=0}^4 {\cal H}_{{\rm int},i}$, 
where 
\begin{eqnarray}
{\cal H}_0 &=& \sum_{p k \sigma} 
\left\{
(\epsilon_{p\kf + k} - \mu ) c^\dagger_{k, p, \sigma} c_{k, p, \sigma}
+ (\epsilon_{3p\kf + k} - \mu ) d^\dagger_{k, p, \sigma} d_{k, p, \sigma}
\right\}, 
\label{eqn:H0D}\\
{\cal H}_{{\rm int},0} &=& 
\frac{1}{L}\sum_{k,k',q} \sum_p \sum_{\sigma,\sigma'} 
\left( \frac{Ua}{2} \delta_{\sigma',-\sigma} -V_2 a \right) 
c^\dagger_{k+q,p,\sigma} c^\dagger_{k'-q,-p,\sigma'}
c_{k',p,\sigma'} c_{k,-p,\sigma} \nonumber \\
&+& \frac{1}{L} \sum_{k,k',q} \sum_p \sum_{\sigma,\sigma'} 
\left( \frac{Ua}{2} \delta_{\sigma',-\sigma} + V_1 a + V_2 a \right) 
c^\dagger_{k+q,p,\sigma} c^\dagger_{k'-q,-p,\sigma'}
c_{k',-p,\sigma'} c_{k,p,\sigma} \nonumber \\
&+& \frac{1}{L} \sum_{k,k',q} \sum_p \sum_{\sigma,\sigma'} 
\left( \frac{Ua}{2} \delta_{\sigma',-\sigma} + V_1 a + V_2 a \right) 
c^\dagger_{k+q,p,\sigma} c^\dagger_{k'-q,p,\sigma'}
c_{k',p,\sigma'} c_{k,p,\sigma}, 
\label{eqn:Hint-0}\\ 
{\cal H}_{{\rm int},1} &=& 
\frac{1}{L} \sum_{k,k',q} \sum_p \sum_{\sigma,\sigma'} 
\left( Ua \delta_{\sigma',-\sigma} - 2 V_2 a \right) 
d^\dagger_{k+q,p,\sigma} c^\dagger_{k'-q,-p,\sigma'}
c_{k',p,\sigma'} c_{k,p,\sigma} \nonumber \\
&+& \frac{1}{L} \sum_{k,k',q} \sum_p \sum_{\sigma,\sigma'} 
\left( Ua \delta_{\sigma',-\sigma} - 2 V_2 a \right) 
c^\dagger_{k+q,p,\sigma} c^\dagger_{k'-q,p,\sigma'}
c_{k',-p,\sigma'} d_{k,p,\sigma}, 
\label{eqn:Hint-1} \\
{\cal H}_{{\rm int},2} &=& 
\frac{1}{L} \sum_{k,k',q} \sum_p \sum_{\sigma,\sigma'} 
\left( Ua \delta_{\sigma',-\sigma} - 2 V_2 a \right) 
c^\dagger_{k+q,p,\sigma} d^\dagger_{k'-q,p,\sigma'}
c_{k',p,\sigma'} d_{k,p,\sigma} \nonumber \\
&+& \frac{1}{L} \sum_{k,k',q} \sum_p \sum_{\sigma,\sigma'} 
\left( Ua \delta_{\sigma',-\sigma} + 2 V_1 a + 2 V_2 a \right) 
c^\dagger_{k+q,p,\sigma} d^\dagger_{k'-q,p,\sigma'}
d_{k',p,\sigma'} c_{k,p,\sigma} \nonumber \\
&+& \frac{1}{L} \sum_{k,k',q} \sum_p \sum_{\sigma,\sigma'} 
\left( \frac{Ua}{2} \delta_{\sigma',-\sigma} -  V_2 a \right) 
\left\{
c^\dagger_{k+q,p,\sigma} c^\dagger_{k'-q,-p,\sigma'}
d_{k',-p,\sigma'} d_{k,p,\sigma} +
h.c. 
\right\} \nonumber \\
&+& \frac{1}{L} \sum_{k,k',q} \sum_p \sum_{\sigma,\sigma'} 
\left( Ua \delta_{\sigma',-\sigma} + 2 V_1 a + 2  V_2 a \right) 
c^\dagger_{k+q,p,\sigma} d^\dagger_{k'-q,-p,\sigma'}
d_{k',-p,\sigma'} c_{k,p,\sigma} \nonumber \\
&+& \frac{1}{L} \sum_{k,k',q} \sum_p \sum_{\sigma,\sigma'} 
\left( \frac{Ua}{2} \delta_{\sigma',-\sigma} -  V_1 a + V_2 a \right) 
\left\{
c^\dagger_{k+q,p,\sigma} c^\dagger_{k'-q,-p,\sigma'}
d_{k',p,\sigma'} d_{k,-p,\sigma} +
h.c. 
\right\} \nonumber \\
&+& \frac{1}{L} \sum_{k,k',q} \sum_p \sum_{\sigma,\sigma'} 
\left( Ua \delta_{\sigma',-\sigma} - 2 V_1 a + 2  V_2 a \right) 
c^\dagger_{k+q,p,\sigma} d^\dagger_{k'-q,-p,\sigma'}
c_{k',p,\sigma'} d_{k,-p,\sigma} \nonumber \\
&+& \frac{1}{L} \sum_{k,k',q} \sum_p \sum_{\sigma,\sigma'} 
\left( \frac{Ua}{2} \delta_{\sigma',-\sigma} -  V_1 a + V_2 a \right) 
\left\{
c^\dagger_{k+q,p,\sigma} c^\dagger_{k'-q,p,\sigma'}
d_{k',-p,\sigma'} d_{k,-p,\sigma} +
h.c. 
\right\} \nonumber \\
&+& \frac{1}{L} \sum_{k,k',q} \sum_p \sum_{\sigma,\sigma'} 
\left( Ua \delta_{\sigma',-\sigma} - 2 V_1 a + 2  V_2 a \right) 
c^\dagger_{k+q,p,\sigma} d^\dagger_{k'-q,p,\sigma'}
c_{k',-p,\sigma'} d_{k,-p,\sigma} \nonumber \\
&+& \frac{1}{L} \sum_{k,k',q} \sum_p \sum_{\sigma,\sigma'} 
\left( Ua \delta_{\sigma',-\sigma} -  2  V_2 a \right) 
c^\dagger_{k+q,p,\sigma} d^\dagger_{k'-q,p,\sigma'}
d_{k',-p,\sigma'} c_{k,-p,\sigma}, 
\label{eqn:Hint-2} \\ 
{\cal H}_{{\rm int},3} &=& 
\frac{1}{L} \sum_{k,k',q} \sum_p \sum_{\sigma,\sigma'} 
\left( Ua \delta_{\sigma',-\sigma} - 2 V_2 a \right) 
c^\dagger_{k+q,p,\sigma} d^\dagger_{k'-q,-p,\sigma'}
d_{k',p,\sigma'} d_{k,p,\sigma} \nonumber \\
&+& \frac{1}{L} \sum_{k,k',q} \sum_p \sum_{\sigma,\sigma'} 
\left( Ua \delta_{\sigma',-\sigma} - 2 V_2 a \right) 
d^\dagger_{k+q,p,\sigma} d^\dagger_{k'-q,p,\sigma'}
d_{k',-p,\sigma'} c_{k,p,\sigma}, 
\label{eqn:Hint-3} \\
{\cal H}_{{\rm int},4} &=& 
\frac{1}{L} \sum_{k,k',q} \sum_p \sum_{\sigma,\sigma'} 
\left( \frac{Ua}{2} \delta_{\sigma',-\sigma} -V_2 a \right) 
d^\dagger_{k+q,p,\sigma} d^\dagger_{k'-q,-p,\sigma'}
d_{k',p,\sigma'} d_{k,-p,\sigma} \nonumber \\
&+& \frac{1}{L} \sum_{k,k',q} \sum_p \sum_{\sigma,\sigma'} 
\left( \frac{Ua}{2} \delta_{\sigma',-\sigma} + V_1 a + V_2 a \right) 
d^\dagger_{k+q,p,\sigma} d^\dagger_{k'-q,-p,\sigma'}
d_{k',-p,\sigma'} d_{k,p,\sigma} \nonumber \\
&+& \frac{1}{L} \sum_{k,k',q} \sum_p \sum_{\sigma,\sigma'} 
\left( \frac{Ua}{2} \delta_{\sigma',-\sigma} + V_1 a + V_2 a \right) 
d^\dagger_{k+q,p,\sigma} d^\dagger_{k'-q,p,\sigma'}
d_{k',p,\sigma'} d_{k,p,\sigma} \point 
\label{eqn:Hint-4} 
\end{eqnarray}
In the above expressions, $L = N_L a$ and 
the matrix elements of the interaction are calculated 
at $K = \pm \kf$ for the lower band and 
at $K = \pm 3 \kf$ for the upper band. 
The elements, 
($Ua/2 \delta_{\sigma',-\sigma} + V_1a + V_2a$), 
($Ua/2 \delta_{\sigma',-\sigma}        - V_2a$) and  
($Ua/2 \delta_{\sigma',-\sigma} - V_1a + V_2a$)
are originated from the interaction processes with the momentum transfer,
$0$, $\pi/(2a)$ and $\pi/a$, respectively.    
Since there is no Umklapp scattering for only the lower band, 
\ie the first term of ${\cal H}_0$ and ${\cal H}_{{\rm int},0}$, 
it is crucial
to take account of the effects of the upper band
in order to describe the insulating state
of the quarter-filled 1-D electron systems.

By averaging the contribution from the upper band,  
the effective Hamiltonian describing the states near $\pm \kf$
is derived as follows.
Representing $c_{k,p,\sigma}$ and $d_{k,p,\sigma}$ in terms of 
   Grassmann variables, 
   the partition function corresponding to 
   eqs.~(\ref{eqn:H0D})-(\ref{eqn:Hint-4})
   is written by   
   $Z = \int {\cal D}[d^*_{k, p, \sigma}d_{k, p, \sigma}]{\cal D}
     [c^*_{k, p, \sigma}c_{k, p, \sigma}]\e^{-S}$,  
 where 
\begin{eqnarray}
S &=& S_{0c} + S_{0d}
          + \sum_{i = 0}^4 S_{{\rm int},i}, 
\label{eqn:S}\\
S_{0c} &=& 
\sum_{p,k,\sigma} \int_0^\beta \d \tau 
c^*_{k, p, \sigma} (\deltau - \mu + \epsilon_{p\kf + k}) c_{k, p, \sigma}, 
\label{eqn:S0c} \\
S_{0d} &=& 
\sum_{p,k,\sigma} \int_0^\beta \d \tau 
d^*_{k, p, \sigma} (\deltau - \mu + \epsilon_{3p\kf + k}) d_{k, p, \sigma}, 
\label{eqn:S0d} \\
S_{{\rm int},i} &=& \int_0^\beta \d \tau {\cal H}_{{\rm int},i},  
\label{eqn:Sint}  
\end{eqnarray}
with $\beta = 1/T$ ($T$: temperature). 
 After integrating with respect to $d_{k, p, \sigma}$, 
    the partition function, $Z$, is given as 
     $Z = Z_{0d} \int {\cal D}[c^*_{k, p, \sigma}c_{k, p, \sigma}]
      \e^{-S_{\rm eff}}$, 
 where 
   $Z_{0d} = \int {\cal D}[d^*_{k, p, \sigma}d_{k, p, \sigma}]
     \e^{-S_{0d}}$, and    
 $S_{\rm eff} = S_{0c} -
         \ln \lan \exp ( - \sum_{i = 0}^4 S_{{\rm int},i} )\ran_d$ 
($\lan A \ran_d \equiv Z^{-1}_{0d} 
\int {\cal D}[d^*_{k, p, \sigma}d_{k, p, \sigma}]
\e^{-S_{0d}} A$).  
The perturbative expansion of $S_{\rm eff}$ leads to 
the effective Hamiltonian.

Up to the second order, 
the normal processes, ${\cal H}_{\rm int, n}$, 
which denote backward scattering and forward scattering, 
are derived as 
\begin{eqnarray} 
{\cal H}_{\rm int, n} &=&
\frac{1}{L} \sum_{k,k',q} \sum_p \sum_{\sigma,\sigma'}
\Big\{
(g_{1\bot}\delta_{\sigma -\sigma'} + g_{1\Vert}\delta_{\sigma \sigma'})
c^\dagger_{k+q, p, \sigma} c^\dagger_{k'-q, -p, \sigma'} c_{k', p, \sigma'} c_{k, -p, \sigma}
\nonumber \\
& & +
(g_{2\bot}\delta_{\sigma -\sigma'} + g_{2\Vert}\delta_{\sigma \sigma'})
c^\dagger_{k+q, p, \sigma} c^\dagger_{k'-q, -p, \sigma'} c_{k', -p, \sigma'} c_{k, p, \sigma} 
\nonumber \\
& & + 
(g_{4\bot}\delta_{\sigma -\sigma'} + g_{4\Vert}\delta_{\sigma \sigma'})
c^\dagger_{k+q, p, \sigma} c^\dagger_{k'-q, p, \sigma'} c_{k', p, \sigma'} c_{k, p, \sigma} 
\Big\} \point
\label{eqn:Hint-n}
\end{eqnarray}
The coupling constants are given as follows,   
\begin{eqnarray}
g_{1 \Vert} 
&=& - V_2 a 
- 4 D_1(- V_2 a)(-V_1 a + V_2 a)  \virg 
\label{eqn:g1-para}\\
g_{1 \bot} 
&=& \frac{Ua}{2} - V_2 a  
- 4 D_1 \left(\frac{Ua}{2} - V_2 a \right) \left(\frac{Ua}{2} - V_1a + V_2 a \right) \virg 
\label{eqn:g1-perp}\\
g_{2 \Vert} 
&=& V_1 a + V_2 a 
- 2 D_1 \left(- V_2 a\right)^2 
- 2 D_1 \left(- V_1 a + V_2 a \right)^2 , 
\label{eqn:g2-para}\\
g_{2 \bot} 
&=& \frac{Ua}{2} + V_1 a + V_2 a  
- 2 D_1 \left(\frac{Ua}{2} - V_2 a \right)^2 
- 2 D_1 \left(\frac{Ua}{2} - V_1 a + V_2 a \right)^2 
              \virg 
\label{eqn:g2-perp} \\
g_{4 \Vert} 
&=& V_1 a + V_2 a  
- 2 D_2 \left(-V_1a + V_2 a\right)^2 \virg 
\label{eqn:g4-para}   \\
g_{4 \bot}
&=& \frac{Ua}{2} + V_1 a + V_2 a  
- 2 D_2 \left(\frac{Ua}{2} - V_1a + V_2 a\right)^2 \virg
\label{eqn:g4-perp}   
\end{eqnarray}
where $D_1 = (8 \pi t a)^{-1} \sqrt{2} \ln (\sqrt{2}+1) 
  \simeq 1.25/(8\pi t a)$ and
$D_2 = (8 \pi t a)^{-1} 2 \sqrt{2}/(\sqrt{2}+1) \sim
1.17/(8 \pi t a)$. 
Note that eqs.~(\ref{eqn:g1-para})-(\ref{eqn:g2-perp}) 
   satisfy the condition, 
$g_{2 \bot} - g_{2 \Vert} +  g_{1 \Vert} =  g_{1 \bot}$, 
   which corresponds to $SU(2)$ symmetry. 
The diagrams which lead to eqs.~(\ref{eqn:g1-para})-(\ref{eqn:g4-perp})
are shown in Fig.~\ref{fig:diagram}, and  
the detailed calculation is given in Appendix A. 
The $8\kf$-Umklapp scattering is obtained from the third order 
expansion shown in Fig.~\ref{fig:umklapp}.   
\begin{figure}
\centerline{\epsfxsize=11.0cm\epsfbox{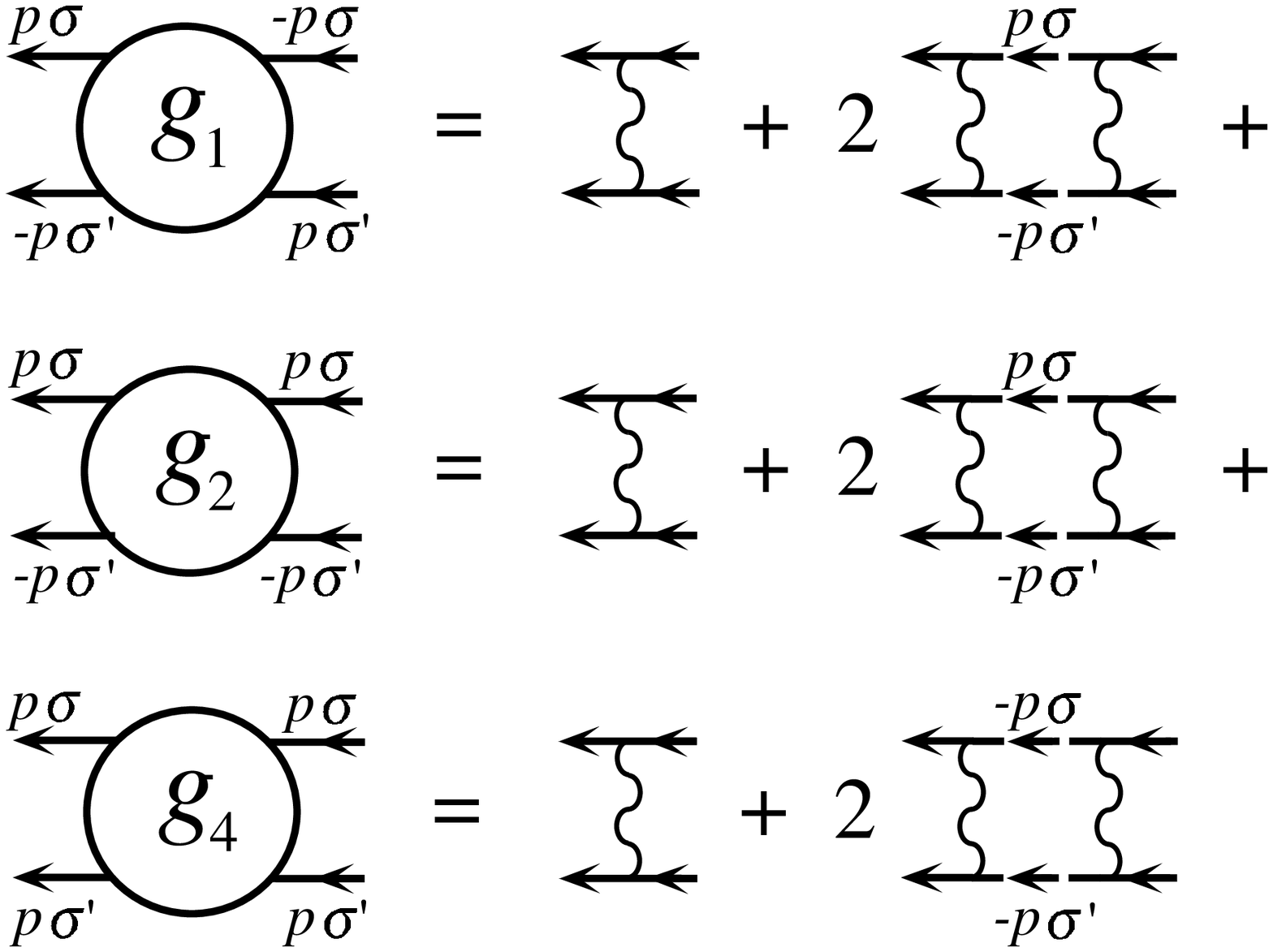}}
\vspace{-1cm}
\caption{ 
Diagrams for the normal scattering. 
Here the dashed and wavy lines express the Green function 
of the upper bands and the interaction,
  respectively. 
}
\label{fig:diagram}
\end{figure}
\begin{figure}
\centerline{\epsfxsize=10.0cm\epsfbox{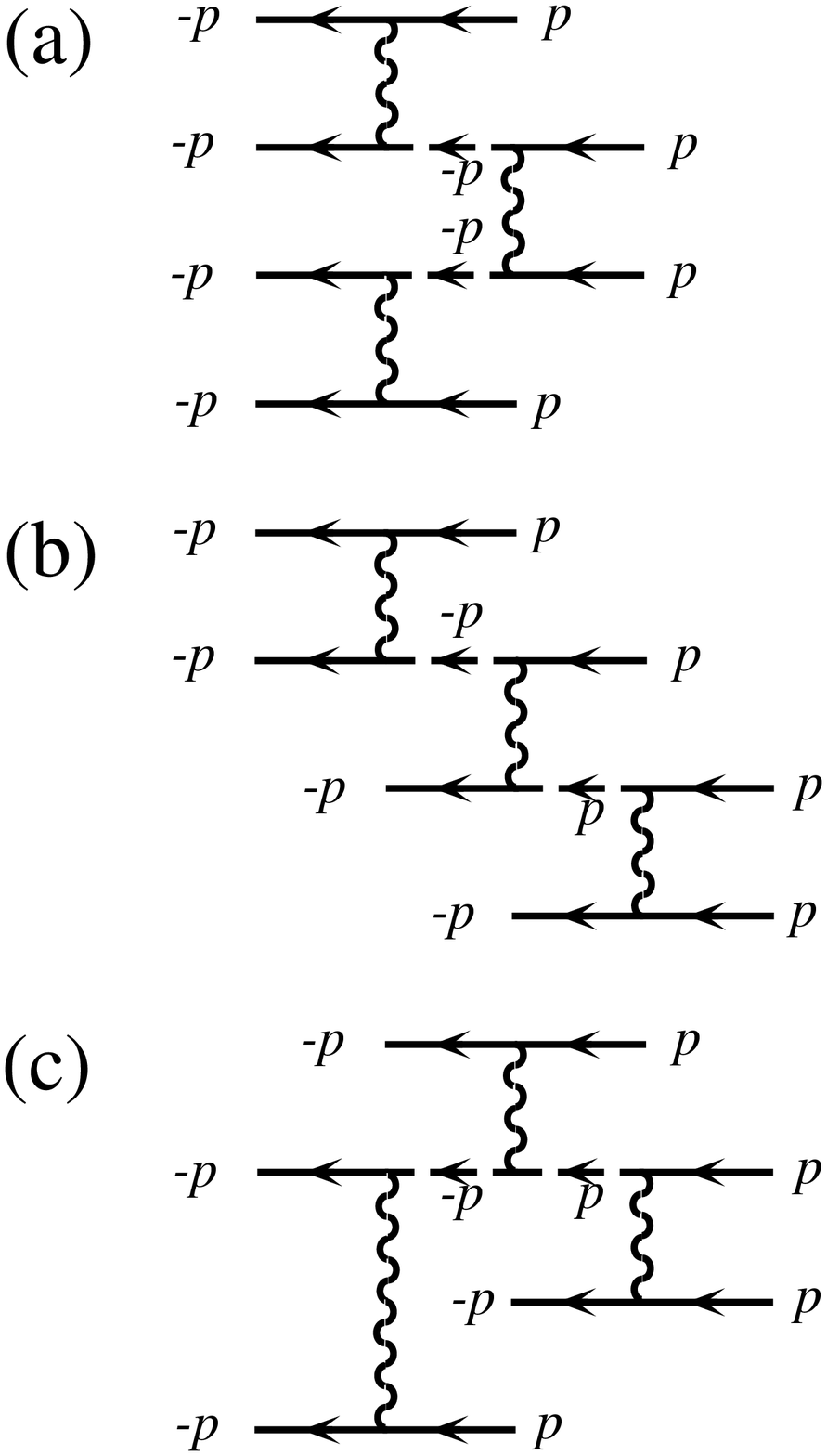}}
\caption{ 
Diagrams for the $8\kf$-Umklapp scattering. 
The dashed and wavy lines express the Green function 
of the upper band and the interaction, respectively. 
}
\label{fig:umklapp}
\end{figure}
 The Hamiltonian for the Umklapp scattering is written as   
\begin{subeqnarray}
{\cal H}_{1/4} &=& 
g'_{1/4} 
\sum_{p} \int \d x (\psi^\dagger_{p,+}\psi_{-p,+})^2 (\psi^\dagger_{p,-}\psi_{-p,-})^2 ,
\label{eqn:H14} \\
g'_{1/4} &=& \frac{1}{2 t^2}
\big\{
(Ua - 2 V_2 a)^2 (Ua - 4 V_1a + V_2 a) 
+ 4 (V_2 a)^2 (Ua - 2 V_1a + 2 V_2 a)
\big\},  
\label{eqn:g'14}
\end{subeqnarray}
where 
$\psi_{p, \sigma} = 1/{\sqrt{L}} \sum_k \e^{\im k x} c_{k,p,
\sigma}$.
Detailed derivation of eqs.~(20a) and (20b) is given in Appendix B. 
Thus, the effective Hamiltonian is written as 
\begin{eqnarray}
{\cal H}_{\rm eff} = 
  {\cal H}_{0 \rm c} + {\cal H}_{\rm{int,n}} + {\cal H}_{1/4}
\virg
\end{eqnarray}
where ${\cal H}_{0 \rm c} = \sum_{p k \sigma} (\epsilon_{p k_F + k} -
\mu) c^\dagger_{k,p,\sigma} c_{k,p,\sigma}$. 
 
\subsection{Bosonization}
We linearize the energy dispersion as 
$\epsilon_{p k_F + k} - \mu \simeq p v_F k$ with $v_F = \sqrt{2} t a$
and utilize the bosonization method. 
The phase variables for the charge (spin) fluctuation,
  $\theta_{\rho}$ and  $\phi_\rho$, 
    ($\theta_{\sigma}$ and $\phi_\sigma$) are defined as
\begin{eqnarray}
\theta_{\rho(\sigma)} &=& 
\sum_{q \ne 0} \frac{\pi \im}{q L}\, \e^{- \alpha |q|/2 -\im q x } 
\sum_{k, p} 
   \left[
       c^\dagger_{k+q, p, +} c_{k, p, +}  +(-) \,
       c^\dagger_{k+q, p, -} c_{k, p, -}
   \right] 
\virg
\label{eqn:theta}\\
\phi_{\rho(\sigma)} &=& 
\sum_{q \ne 0} \frac{\pi \im}{q L} \, \e^{- \alpha |q|/2 -\im q x } 
\sum_{k, p}  p 
   \left[
       c^\dagger_{k+q, p, +} c_{k, p, +}  +(-) \,
       c^\dagger_{k+q, p, -} c_{k, p, -}
   \right] 
\virg
\label{eqn:phi}
\end{eqnarray}
where 
$\alpha^{-1}$ is the ultraviolet cutoff. 
Those fields satisfy the commutation relations, 
$[\theta_{\nu}(x),\phi_{\nu'}(x')] 
= \im \pi \, {\rm sgn}(x-x') \, \delta_{\nu \nu'}$ 
($\nu$, $\nu' = \rho$ or $\sigma$). 
The electron operator is expressed as 
 $\psi_{p, \sigma} = (2 \pi \alpha)^{-1/2}$
     $\exp\{ \im (p/2)(\theta_{\rho} + p \phi_{\rho} 
       + \sigma \theta_{\sigma} + \sigma p \phi_{\sigma})\}$ 
       $\exp(\im \pi \Xi_{p \sigma})$ 
 with 
   $\Xi_{p +} = p/2 \sum_{p'}\N_{p'+}$ and  
  $\Xi_{p -} = p/2 \sum_{p'}\N_{p'-} +\sum_{p'}\N_{p' +}$
 where $\N_{p \sigma} = 
   \int {\rm d} x \psi_{p \sigma}^{\dagger} \psi_{p \sigma}$.  
In terms of these phase variables,
 the effective Hamiltonian is expressed as  
   ${\cal H}_{\rm eff} = {\cal H}_{\rho} + {\cal H}_{\sigma}$,  
   where 
\begin{eqnarray}
{\cal H}_{\rho} &=& \frac{v_{\rho}}{4 \pi} \int \d x
\left\{
\frac{1}{K_{\rho}} (\delx \theta_{\rho})^2 
+ K_{\rho} (\delx \phi_{\rho})^2 
\right\} 
+ \frac{g_{1/4}}{2 (\pi \alpha)^2} \int \d x \cos 4 \theta_{\rho} \virg 
\label{eqn:Hrho} \\ 
{\cal H}_{\sigma} &=& \frac{v_{\sigma}}{4 \pi} \int \d x
\left\{
\frac{1}{K_{\sigma}} (\delx \theta_{\sigma})^2 
+ K_{\sigma} (\delx \phi_{\sigma})^2 
\right\} 
+ \frac{g_{1\bot}}{(\pi \alpha)^2} \int \d x \cos 2 \theta_{\sigma} 
\point
\label{eqn:Hsigma}
\end{eqnarray}
Here
$K_\nu = \sqrt{B_\nu / A_\nu}$ and $v_\nu = \vf \sqrt{B_\nu  A_\nu}$
with $A_{\rho(\sigma)} = 1 + \{g_{4 \Vert}+(-)g_{4 \bot}+g_{2 \Vert}+(-)g_{2
\bot}-g_{1 \Vert}\}/(\pi \vf)$
and $B_{\rho(\sigma)} = 1 + \{g_{4 \Vert}+(-)g_{4 \bot}-g_{2 \Vert}-(+)g_{2
\bot}+g_{1 \Vert}\}/(\pi \vf)$. 
The coupling constant, $g_{1/4}$, is given by
$g_{1/4} = g'_{1/4}/(2 \pi \alpha)^2$.   
In terms of the phase variables, 
the order parameters for the $2\kf$-SDW, $2\kf$-CDW, 
$4\kf$-CDW  and $4\kf$-bond order wave ($4\kf$-BOW)
are written as follows,  
\begin{eqnarray}
O_{2\kf\mbox{-}\rm{SDW}} &=& \sum_{p\sigma} \sigma \e^{-\im 2 p \kf x}
\psi^\dagger_{p, \sigma} \psi_{-p, \sigma} 
\propto \sin(2 \kf x + \theta_{\rho}) \sin \theta_{\sigma} \virg 
\label{eqn:2-SDW}\\
O_{2\kf\mbox{-}\rm{CDW}} &=& \sum_{p\sigma} \e^{-\im 2 p \kf x}
\psi^\dagger_{p, \sigma} \psi_{-p, \sigma} 
\propto \cos(2 \kf x + \theta_{\rho}) \cos \theta_{\sigma} \virg 
\label{eqn:2-CDW} \\
O_{4\kf\mbox{-}\rm{CDW}} &=& \sum_{p} \e^{-\im 4 p \kf x}
\psi^\dagger_{p, +} \psi^\dagger_{p, -}
\psi_{-p, -} \psi_{-p, +} 
\propto \cos(4 \kf x + 2 \theta_{\rho}) \virg
\label{eqn:4-CDW} \\
O_{4\kf\mbox{-}\rm{BOW}} &=& \sum_{p} \e^{-\im 4 p \kf x}
(\im p)
\psi^\dagger_{p, +} \psi^\dagger_{p, -}
\psi_{-p, -} \psi_{-p, +} 
\propto \sin(4 \kf x + 2 \theta_{\rho}) \virg
\label{eqn:4-BOW}
\end{eqnarray}
where the order parameter of 4$\kf$-BOW is given by  
the 4$\kf$-component of $n_{j,+}n_{j+1,-}$. 
Note that 
 the third order perturbation for 
$\ln \langle \exp ( - \sum_{i = 0}^4 S_{{\rm int}, i}) \rangle_d$ 
shown in Fig.~\ref{fig:umklapp} generates 
  the other 8$\kf$-Umklapp scattering term,
$\cos 4 \theta_\rho \cos 2 \theta_\sigma$, 
 which mixes the charge 
and spin fluctuation. 
Since the scaling dimension of this term is smaller than that 
of $\cos 4 \theta_\rho$ or of $\cos 2 \theta_\sigma$, 
we discard the term.   
The effects of the term are discussed in \S 5.

\section{Excitations and Phase Diagram}

We investigate the excitations and the electronic 
 states in the limit of low energy by using the 
 RG method.
At first, we consider the spin degree of freedom. 
By introducing $l = \ln (\alpha'/\alpha)$ as the new length scale 
 $\alpha'$ ($> \alpha$), 
the RG equations for ${\cal H}_\sigma$ are given as follows,\cite{RGS} 
\begin{eqnarray}
\frac{d}{d l} K_{\sigma}(l) &=& - \frac{1}{2} G_{1 \bot}^2(l) K_{\sigma}^2(l) \virg 
\label{eqn:RG-KS}\\
\frac{d}{d l} G_{1 \bot}(l) &=& \left[2 - 2 K_{\sigma}(l)\right] G_{1 \bot}(l) \virg
\label{eqn:RG-G1}
\end{eqnarray}
where the initial conditions are given by 
$K_{\sigma}(0) = K_{\sigma}$ and $G_{1 \bot}(0) = 2 g_{1 \bot} / \pi
v_{\sigma}$. 
The behaviors of $K_{\sigma}(l)$ and $G_{1 \bot}(l)$ are as follows.  
When $g_{1 \bot} \geq 0$ (i.e.,  $U/2 \geq V_2$),   
  the quantities, $G_{1\bot} (l)$ and $K_{\sigma} (l)$ tend to 0 and 1,
    respectively, 
        and the excitation becomes  gapless. 
Then the long-range  correlation functions  are given as 
 $\lan \sin \theta_{\sigma}(x) \sin \theta_{\sigma}(0)\ran
   \sim x^{-1}\ln^{1/2}(x)$ and    
$\lan \cos \theta_{\sigma}(x) \cos \theta_{\sigma}(0)\ran
\sim x^{-1}\ln^{-3/2}(x)$\cite{Giamarchi-Schulz}. 
For $U/2 < V_2$,
the spin gap appears due to $G_{1 \bot} (l) \to - \infty$ and
$K_\sigma (l) \to 0$. 
In this case, the phase, $\theta_\sigma$, is locked as $0$
mod $\pi$ 
at the low energy limit and then    
 $\lan \sin \theta_{\sigma}(x) \sin \theta_{\sigma}(0)\ran$
vanishes exponentially and
$\lan \cos \theta_{\sigma}(x) \cos \theta_{\sigma}(0)\ran$
remains finite at the long distance.
Therefore $2\kf$-SDW does not exist. 
Thus it is found that the spin fluctuation is suppressed by $V_2$. 

\begin{figure}[tbl]
\centerline{\epsfxsize=7.0cm\epsfbox{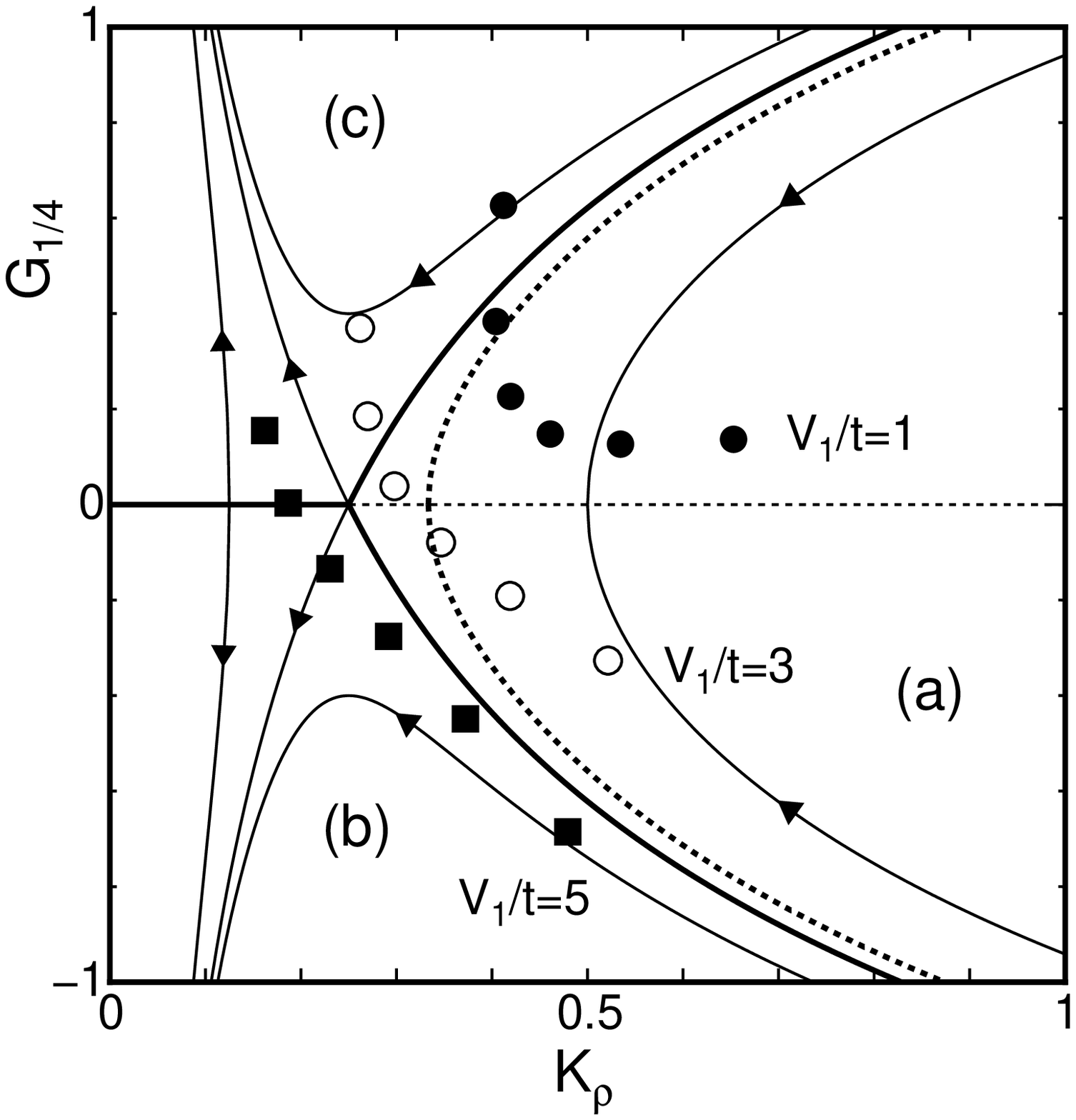}}
\caption{
Scaling flows of eq.~(\ref{eqn:RG-sol})
 on the plane of $K_\rho$ and $G_{1/4}$.\cite{Yoshioka,Yoshioka-II}  
The thick solid curves denote the boundaries 
for metallic regions (a) and insulating regions (b) and (c). 
The fixed point for the thick dotted curve is given by 
 $(K_\rho, G_{1/4}) = (1/3,0)$. 
The closed  circles, open circles and closed squares   
denote the initial values of 
$V_1/t =$ 1, 3 and 5, respectively, 
 with the fixed  $U/t = 6$, 
where respective points from right to left are 
$V_2/t =$ 0, 0.6, 1.2, 1.8, 2.4 and 3.0. 
}
\label{fig:4}
\end{figure}

Next 
we investigate the charge degree of freedom. 
The RG equations corresponding to ${\cal H}_\rho$ 
are written as\cite{Tsuchiizu}   
\begin{eqnarray}
\frac{d}{d l} K_{\rho}(l) &=& - 8 G_{1/4}^2(l) K_{\rho}^2(l) \virg 
\label{eqn:RG-Krho} \\
\frac{d}{d l} G_{1/4}(l) &=& \left[2 - 8 K_{\rho}(l)\right] G_{1/4}(l) \virg
\label{eqn:RG-g14}
\end{eqnarray}  
with the initial conditions given by $K_{\rho}(0) = K_{\rho}$ 
and $G_{1/4}(0) = g_{1/4}/(2 \pi v_{\rho})$. 
Correlation functions are calculated from the solutions of 
  eqs.~(\ref{eqn:RG-Krho}) and (\ref{eqn:RG-g14}) as, 
$ \lan \sin \theta_\rho(x) \sin \theta_\rho(0) \ran 
=  \lan \cos \theta_\rho(x) \cos \theta_\rho(0) \ran
\propto \exp \{
- \int_0^{\ln x/\alpha} \d l K_\rho(l)
\} $,   
$  \lan \cos 2 \theta_\rho(x) \cos 2 \theta_\rho(0) \ran
\propto \exp \{
- \int_0^{\ln x/\alpha} \d l [ 4 K_\rho(l) + 2 G_{1/4}(l) ]
\} $ and 
$  \lan \sin 2 \theta_\rho(x) \sin 2 \theta_\rho(0) \ran
\propto \exp \{
- \int_0^{\ln x/\alpha} \d l [ 4 K_\rho(l) - 2 G_{1/4}(l) ]
\} $.
\cite{CF}
Equations (\ref{eqn:RG-Krho}) and (\ref{eqn:RG-g14})
are calculated analytically as\cite{Yoshioka}
\begin{eqnarray}
G^2_{1/4}(l) - 1/(2K_{\rho}(l)) - 2 \ln K_{\rho}(l) = A \virg
\label{eqn:RG-sol}
\end{eqnarray}
where $A$ is a numerical constant determined by the initial conditions.
Equation (\ref{eqn:RG-sol}) is obtained by rewriting
   eq.~(\ref{eqn:RG-Krho}) as
   $G^2_{1/4}(l) = (1/8) d K_\rho^{-1} (l)/ d l$ and 
   $K_{\rho} (l) G^2_{1/4}(l) = - (1/8) d \ln K_\rho (l) / d l$, 
   which are substituted into eq.~(\ref{eqn:RG-g14}).   
By choosing some parameters, 
we show eq.~(\ref{eqn:RG-sol}) on the plane of $K_\rho$ and $G_{1/4}$ 
in Fig.~\ref{fig:4}. 
In the region (a), 
  the non-linear $G_{1/4}$ term becomes irrelevant leading to  metallic state.
Here the correlation functions with  long distance are given by 
  $\lan \sin \theta_{\rho}(x) \sin \theta_{\rho}(0)\ran
   = \lan \cos \theta_{\rho}(x) \cos \theta_{\rho}(0)\ran
    \sim x^{- K_{\rho}(\infty)}$
and $\lan \cos 2 \theta_{\rho}(x) \cos 2 \theta_{\rho}(0)\ran
   \sim \lan \sin 2 \theta_{\rho}(x) \sin 2 \theta_{\rho}(0)\ran
   \sim x^{- 4 K_{\rho}(\infty)}$. 
In particular, the initial values located between the thick solid curve and 
the thick dotted one are renormalized to $1/4 < K_\rho (\infty) < 1/3$ and 
$G_{1/4}(\infty) = 0$.  
Both regions (b) and (c) corresponds to the strong coupling regime 
with insulating states, in which $|G_{1/4}(l)| \to \infty$ and 
$K_\rho(l) \to 0$.    
In the region (b) where $G_{1/4}(l) \to -\infty$, 
the correlation functions, 
$\lan \cos 2 \theta_{\rho}(x) \cos 2 \theta_{\rho}(0)\ran$,  
$\lan \cos \theta_{\rho}(x) \cos \theta_{\rho}(0)\ran$, and 
$\lan \sin \theta_{\rho}(x) \sin \theta_{\rho}(0)\ran$
remain finite at the long distance, 
whereas $\lan \sin 2 \theta_{\rho}(x) \sin 2 \theta_{\rho}(0)\ran$
decays exponentially.    
Then both $2\kf$-density waves and $4\kf$-CDW states are stabilized.
On the other hand, in the region (c) where $G_{1/4}(l) \to \infty$, 
one finds the finite value of
$\lan \sin 2 \theta_{\rho}(x) \sin 2 \theta_{\rho}(0)\ran$, 
$\lan \cos \theta_{\rho}(x) \cos \theta_{\rho}(0)\ran$ and 
$\lan \sin \theta_{\rho}(x) \sin \theta_{\rho}(0)\ran$
but the exponential decay for 
$\lan \cos 2 \theta_{\rho}(x) \cos 2 \theta_{\rho}(0)\ran$. 
Therefore $4\kf$-BOW and $2\kf$-density waves are stabilized and 
$4\kf$-CDW is completely suppressed. 
In Fig.~\ref{fig:4}, the closed  circles, open circles and closed
  squares 
denote the initial values for $V_1/t =$ 1, 3 and 5, respectively, 
 with the fixed  $U/t = 6$, where respective points from right to left  are   
$V_2/t =$ 0, 0.6, 1.2, 1.8, 2.4 and 3.0. 
Here $\alpha \simeq 2a/\pi$ is used.\cite{Yoshioka} 
Since increase of $V_2$ with fixed $U$ and $V_1$ 
moves ($K_\rho$, $G_{1/4}$) to 
the region (c),  
it is found that the next-nearest-neighbor repulsion $V_2$
stabilizes $4\kf$-BOW states.

\begin{figure}[tbl]
\centerline{\epsfxsize=7.0cm\epsfbox{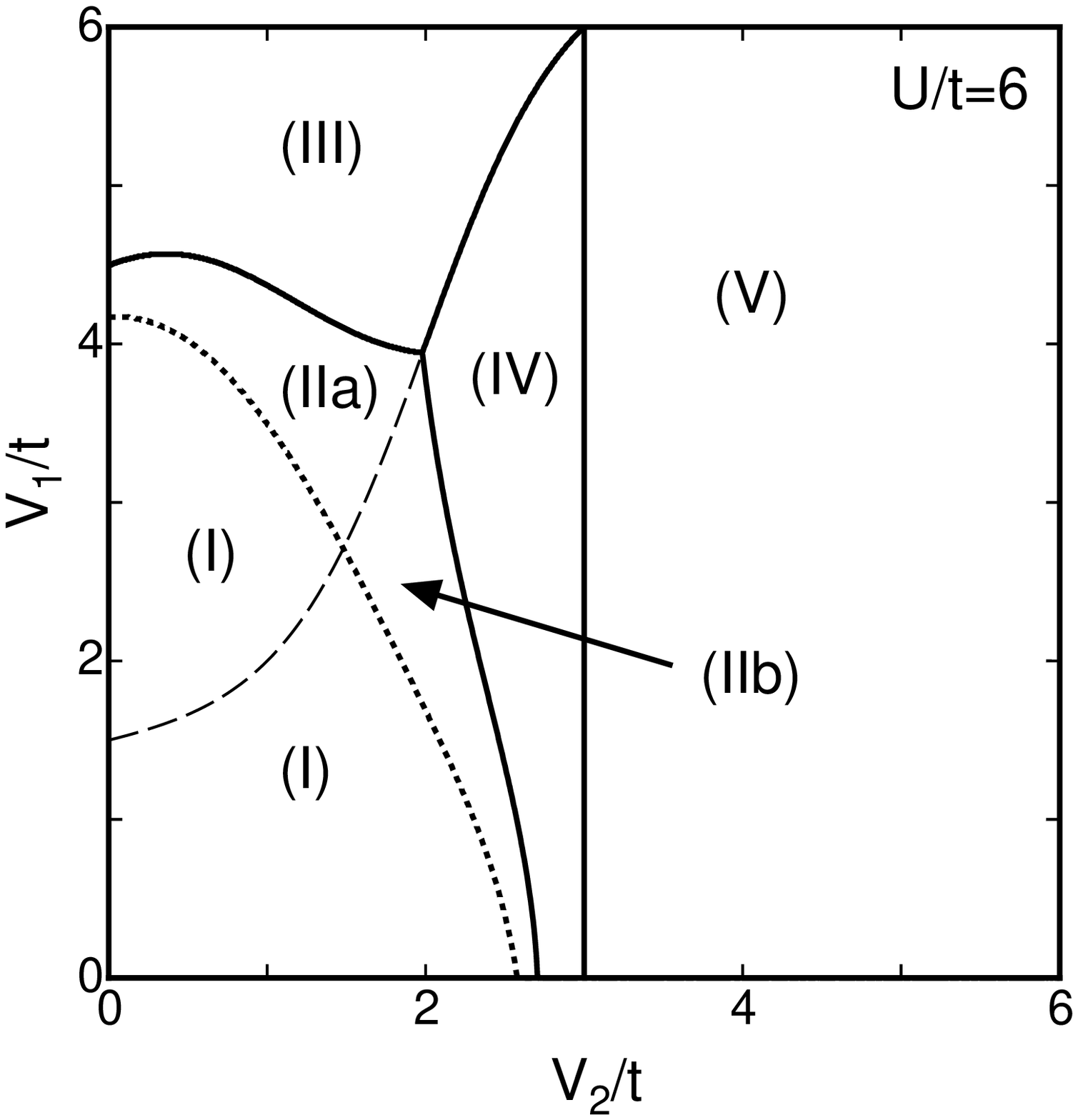}}
\caption{
The phase diagram on the plane of $V_1/t$ and $V_2/t$ 
   at $U/t = 6$ where the thin dashed curve expresses $g_{1/4} = 0$.  
The states of respective regions are given by 
   2$\kf$-SDW metal (I),
   4$\kf$-CDW metal (IIa),
   4$\kf$-BOW metal (IIb),
   4$\kf$-CDW insulator (III),
   4$\kf$-BOW insulator (IV),
   and 2$\kf$-CDW + 4$\kf$-BOW insulator (V).
In the region (III) and (IV),  
   2$\kf$-SDW exists as the dominant fluctuation.  
}
\label{fig:5}
\end{figure}

By combining above consideration on the charge and spin degrees of 
   freedom, we obtain the phase diagram, which is shown on the plane 
   of $V_1/t$ and $V_2/t$ with fixed $U/t = 6$ in Fig.~\ref{fig:5}. 
Here the most dominant state of the respective region is as follows;  
   2$\kf$-SDW metal (I),
   4$\kf$-CDW metal (IIa),
   4$\kf$-BOW metal (IIb),
   4$\kf$-CDW insulator (III),
   4$\kf$-BOW insulator (IV),
   and 2$\kf$-CDW + 4$\kf$-BOW insulator (V).
%
The region (I) with small $V_1/t$ and $V_2/t$, 
   which exhibits the gapless excitation for both 
   charge and spin,
   denotes the metallic state with $2\kf$-SDW as  
   the most dominant fluctuation.
In the region (III), the insulating ordered state of $4\kf$-CDW 
   appears, where $2\kf$-SDW exists as the most dominant fluctuation. 
This comes from the fact that 
   the parameters located in the metallic region (a) in Fig.~\ref{fig:4}
   move to the insulating region (b) 
   with increase of $V_1$ and fixed small $V_2$, 
   whereas the spin excitation remains gapless. 
Note that between the insulating state of $4\kf$-CDW and 
   the metallic state of $2\kf$-SDW, the metallic state with 
   the most dominant fluctuation of $4\kf$-CDW exists (region (IIa)). 
Such a region is obtained for 
   $1/4 < K_{\rho}(\infty) < 1/3$ and $g_{1/4} <0$ 
   (see Fig.~\ref{fig:4}). 
With increasing $V_2/t$, the parameters move from the region (b) to the
   region (c) in Fig.~\ref{fig:4}.
In this case, the 4$\kf$-CDW state disappears and 
   the ordered state of $4\kf$-BOW appears (region (IV)), where
   the $2\kf$-SDW state remains as the most dominant fluctuation   
   as long as the spin excitation is gapless.
The metallic state of $2\kf$-SDW in the region (I) turns into 
   the insulating $4\kf$-BOW state (region (IV))  when $V_2$ 
   increases with fixed small $V_1$ 
   as seen from the closed circles with 
   $V_1/t = 1$  in Fig.~\ref{fig:4}. 
With increasing $V_2$ further, 
   the spin gap appears for $U < 2 V_2$ and the insulating 
   $2\kf$-CDW + $4\kf$-BOW state is
   realized in the region (V) since the correlation functions of 
   $2\kf$-SDW and of $4\kf$-CDW are exponentially decayed 
   at the long distance.   
In contrast to the region (IIa),
   there exists the metallic 4$\kf$-BOW state (region (IIb)) 
   corresponding to $1/4 < K_{\rho}(\infty) < 1/3$ and $g_{1/4} > 0$
   between the region (I) and (IV).

  
\begin{figure}[tbl]
\centerline{\epsfxsize=7.0cm\epsfbox{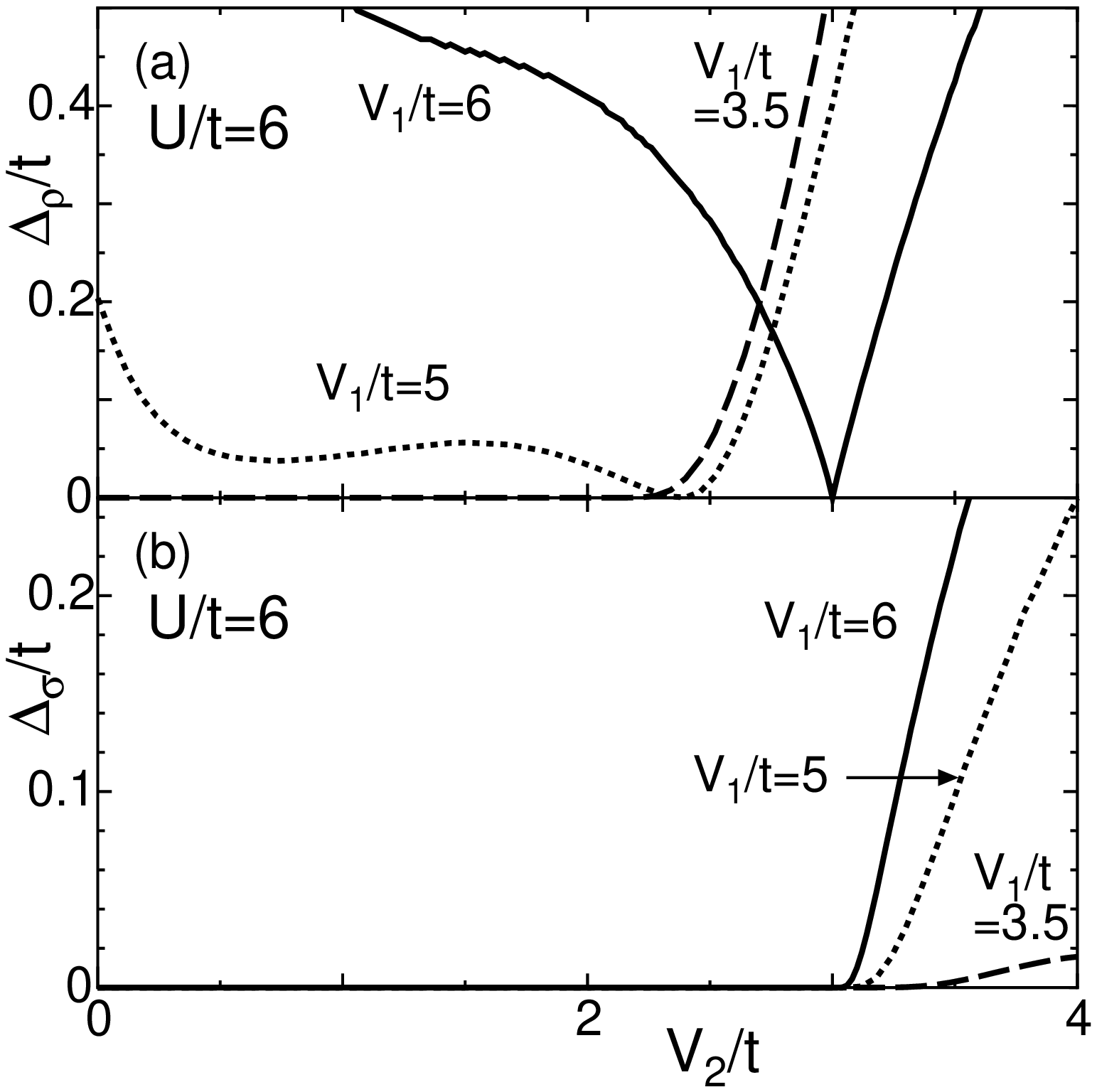}}
\caption{
The charge gap $\Delta_\rho$ (a) and the spin gap $\Delta_\sigma$
(b) as a function of $V_2$ at $U/t=6$ with choices of 
$V_1/t = 6$ (solid curve), $V_1/t = 5$ (dotted curve) and 
$V_1/t=3.5$ (dashed curve). 
}
\label{fig:6}
\end{figure}
 Figure \ref{fig:6} shows the charge gap $\Delta_\rho$ (a) and the spin gap 
$\Delta_\sigma$ (b) as a function of $V_2$ with 
choices of $V_1/t = 6$, $V_1/t = 5$ and $V_1/t = 3.5$ for $U/t=6$. 
By using the solution of the RG equations, (\ref{eqn:RG-KS})-(\ref{eqn:RG-g14}), 
the quantities $\Delta_\rho$ and $\Delta_\sigma$ are determined by 
the following condition, 
$\Delta_\rho = v_\rho \alpha^{-1} \exp (- l_\rho)$ 
with $|G_{1/4}(l_\rho)| = 1$,\cite{Tsuchiizu} and 
$\Delta_\sigma = v_\sigma \alpha^{-1} \exp (- l_\sigma)$ 
with $G_{1 \bot}(l_\sigma) = -2$. 
Note that 
the spin gap determined by such a method 
shows quantitatively good agreement with 
that obtained from the exact solution\cite{Bahder} 
for $\Delta_\sigma /t< 1$ 
in case of negative $U$ Hubbard model.  
For $V_1/t =6 $ and 5, 
   the charge gap is finite except for the boundary between 
   the region (III) and the region (IV) in Fig.~5.
In the regions (IV) and (V), 
$\Delta_\rho$ increases monotonously with $V_2$.
In the region (III), $\Delta_\rho$ for $V_1/t = 5$ 
   does not show a monotonous behavior as a function of $V_2$,  
   whereas $\Delta_\rho$ for $V_1/t = 6$ decreases 
   with increase of $V_2$. 
Such a behavior of $\Delta_\rho$ for $V_1/t = 5$
   may come from the fact that 
   the boundary between the region (IIa) and the region (III)
   is not a monotonous function of $V_2$.  
As mentioned above, 
the condition for appearance of the spin gap 
is given by $U < 2 V_2$ and then independent of $V_1$. 
However, as seen in Fig.~\ref{fig:6}(b), 
the magnitude of the spin gap 
does depend on the value of $V_1$,  
which enhances the spin gap.

Here we note the case of small $U$.  
The metallic state is realized for all regions corresponding 
   to Fig.~\ref{fig:5}, 
\ie the region for $V_1 < U$ and $V_2 < U$.  
In this case,  
   the new metallic state appears for $U < 2 V_2$
   in which the spin gap exists as shown in just below 
   eq.~(\ref{eqn:RG-G1}).  
In this metallic state, the most dominant fluctuation is 
$2\kf$-CDW for $K_\rho (\infty) < 1$.

Finally, we consider correspondence between the present state
and the coexistent state found by the mean-field theory. 
As discussed in the previous paper,\cite{Yoshioka} 
   the insulating $4\kf$-CDW states in the region (III) 
   corresponds to the coexistent
   state of $4\kf$-CDW and $2\kf$-SDW \cite{Seo-Fukuyama}
   because the most dominant fluctuation of $2\kf$-SDW 
   exists in the region (III).  
In the insulating 4$\kf$-BOW state (IV), 
   the fluctuations of both 2$\kf$-CDW and 2$\kf$-SDW develop, 
   whereas the 4$\kf$-CDW state is completely suppressed. 
Therefore, the state of the region (IV) may correspond to 
   the coexistent states
   of two kinds of 2$\kf$-density waves 
   \cite{Kobayashi-Ogata-Yonemitsu}
   without $4\kf$-CDW.
   \cite{Tomio-Suzumura}

\section{Spin Susceptibility}

\begin{figure}
\centerline{\epsfxsize=7.0cm\epsfbox{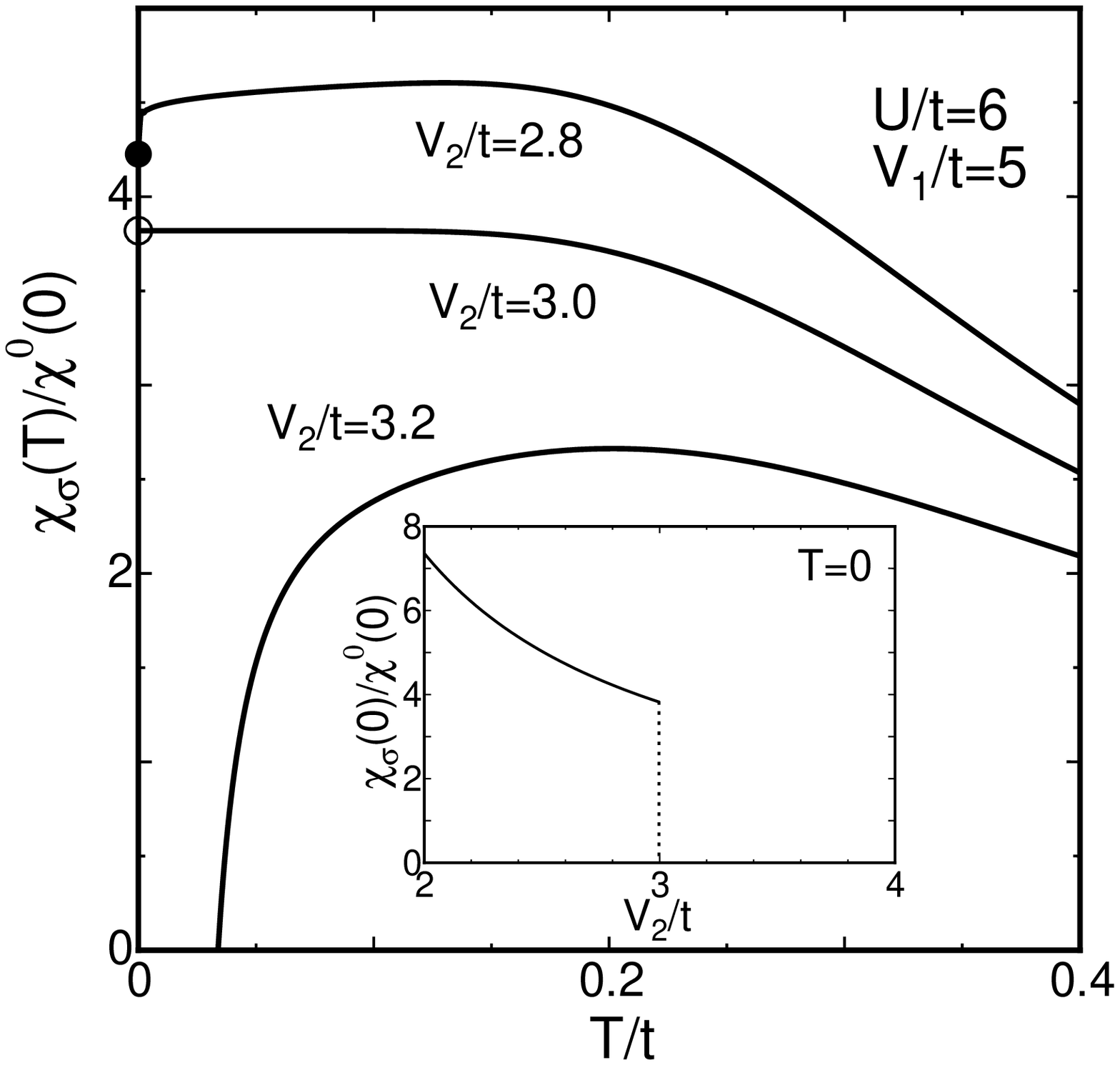}}
\caption{ 
Normalized spin susceptibility, $\chi_\sigma(T)/\chi^0 (0)$, 
where $\chi^0 (0)$ is the susceptibility  
in the absence of the interaction at $T=0$, 
 as a function of $T/t$ for the several choice of $V_2$
with fixed $U/t = 6$ and $V_1/t =5$. 
The black and white circles denote the values of $T=0$ 
for $U > 2 V_2$ and $U = 2V_2$, respectively.
Inset : Spin susceptibility at $T = 0$ as a function of $V_2/t$
with fixed $U/t = 6$ and $V_1/t =5$. 
}
\label{fig:chis}
\end{figure}

In this section, the spin susceptibility of the present
system is investigated. 
We utilize the formula developed in ref.~\citen{Nelisse}, 
which is effective at relatively low temperatures
since it is based on the RG analysis using the linearized band. 
From eqs.~(\ref{eqn:RG-KS}) and (\ref{eqn:RG-G1}), 
one obtains 

\begin{eqnarray}
g_{1\bot}(l) &=& \frac{g_{1\bot}}{1 + 2 g_{1 \bot}(\pi v_\sigma^0)^{-1}l} \virg
\label{eqn:g1-s}
\end{eqnarray}
where $g_{1\bot}/ (\pi v_\sigma) \ll 1$ and 
$v_\sigma^0 = \vf \left\{ 1 + (g_{4 \Vert}-g_{4\bot})(\pi \vf)^{-1} \right\}$. 
By using eq.~(\ref{eqn:g1-s}), 
the spin susceptibility, $\chi_\sigma(T)$, is calculated as\cite{Nelisse}
\begin{eqnarray}
\chi_\sigma (T) &=& \frac{1}{2} 
\frac{\chi_p^0(T)}{1 - \{g_{4 \bot} - g_{4 \Vert} + g_{1 \bot}(l)\} \chi_p^0(T)} \virg 
\label{eqn:chis}\\
\chi_p^0(T) &=& - \frac{2}{L} \sum_k  \lim_{q \to 0} 
\frac{f(\epsilon_{p k_F + k}) - f(\epsilon_{p k_F + k+q})}
{\epsilon_{p k_F + k} - \epsilon_{p k_F + k + q}},
\label{eqn:chi0}
\end{eqnarray}  
where $f(\epsilon) = 1/(\e^{(\epsilon - \mu)/T} + 1)$ and $l = \ln (\vf
\al^{-1}/T)$.  
In Fig.~\ref{fig:chis}, 
$\chi_\sigma (T) / \chi^0 (0)$ is shown where
$\chi^0 (0)$ denotes the susceptibility in the  
absence of the interaction at $T=0$. 
At high temperatures, the spin susceptibility increases 
with decreasing temperature. 
In the limit of zero temperature, 
the susceptibility approaches to a constant with infinite slope for $U > 2 
V_2$ as seen in Hubbard model.\cite{Nelisse}
Since it is due to logarithmic temperature dependence of $g_{1 \bot}(l)$, 
such an anomaly is not observed for $U = 2 V_2$. 
In case of the strong coupling, \ie $U < 2 V_2$, the susceptibility vanishes near 
the temperature, $T_{SG} = \vf \al^{-1} \exp\{\pi v_\sigma^0/(2 g_{1 \bot})\}$, 
which originates from the fact that we use the solution of 
the weak coupling expansion,
eq.~(\ref{eqn:g1-s}). 
It is expected that the susceptibility 
for $T \ll T_{SG}$ 
behaves as
$(\Delta_\sigma/T)^{1/2} \exp\{-\Delta_\sigma/T\}$\cite{Luther}  
where $\Delta_\sigma$ is the spin gap.  
The inset shows $\chi_\sigma (0) / \chi^0 (0)$
as a function of $V_2/t$
with fixed $U/t = 6$ and $V_1/t =5$. 
The interaction  $V_2$ suppresses the spin fluctuation monotonously  
even for the region without the spin gap. 
On the other hand, $V_1$ enhances $\chi_\sigma (0) / \chi^0 (0)$. 
These are due to the effect from the upper band 
because $\chi_\sigma (0) / \chi^0 (0) = \left\{ 1 + (g_{4 \Vert}-g_{4\bot})(\pi
\vf)^{-1} \right\}^{-1}$
with $g_{4 \Vert}-g_{4\bot} = - Ua /2 
\{
1 - 2 D_2 ( Ua/2 - 2 V_1 a + 2 V_2 a)
\}$.
Note that the spin susceptibility in case of $V_1 = V_2 = 0$ is
different from that derived in ref.~\citen{Nelisse}
because our formula includes the virtual process to the upper band. 
Spin susceptibilities 
  both with and without the virtual processes 
   are compared with that 
   derived from the exact solution\cite{Shiba} in Appendix C.    
 
\section{Summary}

We investigated
the electronic states and excitations of the 
one-dimensional electron system at quarter-filling
with on-site ($U$), 
nearest-neighbor ($V_1$)  and 
next-nearest-neighbor ($V_2$) repulsive interactions. 
By taking account of the effects 
of the upper band in a perturbative way, 
the effective Hamiltonian has been obtained 
in terms of the phase variables 
and analyzed by the renormalization group method. 
  
The following roles of $V_2$ have been found. 
The spin fluctuation is suppressed with increasing $V_2$ and 
the spin gap is obtained for $V_2 > U/2$.
Such a characteristic behavior can be seen in the spin susceptibility,
$\chi_\sigma(T)$, where  
$\chi_\sigma(0)$ decreases with increasing $V_2$ and 
vanishes for $V_2 > U/2$. 
Note that the decrease of  $\chi_\sigma(0)$ by $V_2$ for $U/2 > V_2$
   is due to the effects of the upper band.  
In addition, the interaction suppresses the $4\kf$-CDW state and 
   stabilizes the $4\kf$-BOW state.  
As a result, the insulating state of $4\kf$-CDW realized for 
large $U$ and $V_1$ changes into the insulating state of $4\kf$-BOW
with increasing $V_2$. 
The insulating $4\kf$-BOW state moves to 
the insulating states of $4\kf$-BOW + $2\kf$-CDW 
with increasing $V_2$ further. 
Thus the charge distribution changes  
from $(\bigcirc \circ \bigcirc \circ)$ of $4\kf$-CDW to 
$(\bigcirc \bigcirc \circ \circ)$ of $2\kf$-CDW 
with increase of $V_2$ where
$\bigcirc$ ($\circ$) denotes the rich (poor) charge 
at the respective site.  
The excitation gaps were determined as a function of $V_2$
with several choices of $V_1$ at $U/t=6$ 
from the solutions of the RG equations. 
In the states of insulating $4\kf$-BOW and $4\kf$-BOW + $2\kf$-CDW, 
the charge gap increases as a function of $V_2$, 
whereas the gap is not a monotonous function of 
$V_2$ in the insulating $4\kf$-CDW state 
when parameters are located 
near the boundary 
of metal insulator transition. 
The spin gap appears independently of the value of $V_1$. 
However, the magnitude of the spin gap 
does depend on $V_1$, which enhances the spin gap.   
For the case of weak on-site repulsion, 
the ground state for $V_1 < U$ is metallic. 
When $V_2$ is added to the state, 
the metallic state with spin gap appears, 
where the dominant fluctuation is given 
by the $2\kf$-CDW. 

In the present treatment, we have neglected  
the other $8\kf$-Umklapp scattering term, 
$\cos 4 \theta_\rho \cos 2 \theta_\sigma$,
   which mixes the charge and spin degrees of freedom and has 
the scaling dimension less than 
that of $\cos 2 \theta_\sigma$ or of $\cos 4 \theta_\rho$. 
The importance of such a term  
has been pointed out 
for half-filling extended Hubbard model with $U + V_1$,\cite{Cannon,Voit,Nakamura}  
in which a term $\cos 2 \theta_\rho \cos 2 \theta_\sigma$ exists 
besides $\cos 2 \theta_\rho$ and $\cos 2 \theta_\sigma$
  (the definitions of the phases and 
   the interaction correspond to ours).
In the half-filling model with $U+V_1$, the coefficient of $\cos 2 \theta_\rho$ 
and of $\cos 2 \theta_\sigma$ vanishes at $U=2V_1$, 
whereas the term $\cos 2 \theta_\rho \cos 2 \theta_\sigma$ is finite. 
Thus  the new phase appears near $U \simeq 2 V_1$.\cite{Nakamura} 
In the present model, 
the coefficient of $\cos 4 \theta_\rho$ and that of $\cos 2
\theta_\sigma$
become zero simultaneously
at the point, $U = V_1 = 2 V_2$. 
In this point, 
the coefficient of $\cos 4 \theta_\rho \cos 2 \theta_\sigma$ 
also vanishes (see Appendix B). 
Therefore, 
it is expected that such a term does not change  
the phase diagram qualitatively, but may
result in  only a quantitative change. 

The appearance of the spin gap and the variation 
of the charge order with increasing $V_2$ 
are of interest and 
may be observed by the increase of the pressure  
along the conducting chain.\cite{Kagoshima-II}

\vspace{1em}
\section*{Acknowledgment}
The authors would like to thank H. Fukuyama, T. Itakura, H. Seo, 
Y. Tomio 
for stimulating discussions. 
 This work was partly supported by 
 a Grant-in-Aid 
 for Scientific  Research  from the Ministry of Education, 
Science, Sports and Culture, (No.11740196)
 and by  
a Grant-in-Aid 
 for Scientific  Research  from the Ministry of Education, 
Science, Sports and Culture, (No.09640429). 

\appendix
\section{Derivation of the normal processes} 

From Fig.~2, the matrix elements are given as
\begin{eqnarray}
\label{g1}
g_{1,\sigma \sigma'} &=& X_{2\kf,\sigma \sigma'} 
-4 \frac{1}{\beta L} \sum_{-\pi/(4a)<k\leq\pi/4a} \sum_{\en}
X_{2\kf,\sigma \sigma'} X_{4\kf,\sigma \sigma'}
G_p(\im \en,k) G_{-p}(-\im \en,-k), \\
\label{g2}
g_{2,\sigma \sigma'} &=& X_{0,\sigma \sigma'} 
-2 \frac{1}{\beta L} \sum_{-\pi/(4a)<k\leq\pi/4a} \sum_{\en}
\left\{ X_{2\kf,\sigma \sigma'}^2 + X_{4\kf,\sigma \sigma'}^2 \right\} 
G_p(\im \en,k) G_{-p}(-\im \en,-k), \\ 
\label{g4}
g_{4,\sigma \sigma'} &=& X_{0,\sigma \sigma'} 
-2 \frac{1}{\beta L} \sum_{-\pi/(4a)<k\leq\pi/4a} \sum_{\en}
X_{4\kf,\sigma \sigma'}^2 
G_{-p}(\im \en,k) G_{-p}(-\im \en,-k),
\end{eqnarray}
where $g_{i,\sigma \sigma} = g_{i \Vert}$ 
and $g_{i,\sigma -\sigma} = g_{i \bot}$ ($i=$1,2,4). 
The quantities, $X_{0,\sigma \sigma'}$, $X_{2\kf,\sigma \sigma'}$ and 
$X_{4\kf,\sigma \sigma'}$ are the matrix elements of the interaction, 
$X_{0,\sigma \sigma'} = Ua/2 \delta_{\sigma,-\sigma'} + V_1 a + V_2 a$, 
$X_{2\kf,\sigma \sigma'} = Ua/2 \delta_{\sigma,-\sigma'} - V_2 a$ and 
$X_{4\kf,\sigma \sigma'} = Ua/2 \delta_{\sigma,-\sigma'} - V_1 a + V_2 a$, respectively. 
The green function of the upper band is given by 
$G_p(\im \en, k) = 1/(\im \en - \epsilon_{3p \kf + k} + \mu)$.
The particle-particle propagaters are calculated as,
\begin{eqnarray}
\label{d1}
D_1 &=& \frac{1}{\beta L} \sum_{-\pi/(4a)<k\leq\pi/4a} \sum_{\en}
 G_p(\im \en,k )G_{-p}(-\im \en,-k ) 
\nonumber \\
&=& \frac{1}{L} \sum_{-\pi/(4a)<k\leq\pi/4a} 
\frac{1}{2(\epsilon_{3p\kf + k} - \mu)}
\left\{ 1 - 2 f(\epsilon_{3p\kf + k} - \mu)\right\} \nonumber \\
&\simeq& \frac{1}{8 \pi t a} \int_0^{\pi/2} \d y \frac{1}{\sin y + 1/\sqrt{2}}
= \frac{\sqrt{2}}{8\pi ta} \ln (\sqrt{2}+1) \virg \\
\label{d2}
D_2 &=& \frac{1}{\beta L} \sum_{-\pi/(4a)<k\leq\pi/4a} \sum_{\en}
 G_p(\im \en,k )G_{p}(-\im \en,-k ) 
\nonumber \\
&=& \frac{1}{L} \sum_{-\pi/(4a)<k\leq\pi/4a} 
\frac{1}{\epsilon_{3p\kf + k} + \epsilon_{3p\kf - k} - 2\mu}
\left\{ 1 - f(\epsilon_{3p\kf + k} - \mu) - f(\epsilon_{3p\kf - k} - \mu)\right\} \nonumber \\
&\simeq& \frac{1}{4 \pi t a} \int_0^{\pi/2} \d y 
\frac{1}{\sin y + \cos y + \sqrt{2}}
= \frac{1}{8\pi ta} \frac{2 \sqrt{2}}{\sqrt{2}+1} \virg  
\end{eqnarray}
where $f(\epsilon) = 1/(\e^{\beta \epsilon} + 1)$. 
Here $f(\epsilon_{3p\kf \pm k} - \mu)$ is treated as zero
because the energy scale we consider is smaller than $-\mu = \sqrt{2} t$.
Equations (\ref{g1})-(\ref{d2}) give rise to 
eqs.~(\ref{eqn:g1-para})-(\ref{eqn:g4-perp}).

\section{Derivation of Umklapp scattering}
As shown in Fig.~3, 
the $8\kf$-Umklapp scattering consists of the interaction processes,  
in which the right moving four electrons are scattered into the left 
moving states and vice versa. 
Such processes are derived from 
$\lan S_{{\rm int}, 1}^2 S_{{\rm int}, 2} \ran_d / 2$
among the third order perturbation. 
The processes in Fig.~3(a), (b) and (c) 
are written as $S_a$, $S_b$ and $S_c$, respectively,    
where
\begin{eqnarray}
S_a &=& \frac{1}{2L^3}\sum_p 
\sum_{\sigma_1,\sigma_1'} \sum_{k_1,k_1',q_1} \int_0^\beta \d \tau_1
2 X_{2\kf, \sigma_1 \sigma_1'}
\sum_{\sigma_2,\sigma_2'} \sum_{k_2,k_2',q_2} \int_0^\beta \d \tau_2
2 X_{2\kf, \sigma_2 \sigma_2'}
\sum_{\sigma_3,\sigma_3'} \sum_{k_3,k_3',q_3} \int_0^\beta \d \tau_1
X_{4\kf, \sigma_3 \sigma_3'} \nonumber \\
&\times&
\Big\{
\langle
c^*_{k_1+q_1,-p,\sigma_1}(\tau_1)c^*_{k_1'-q_1,-p,\sigma_1'}(\tau_1)
c_{k_1',p,\sigma_1'}(\tau_1)d_{k_1,-p,\sigma_1}(\tau_1) \nonumber \\
&\times&
c^*_{k_2+q_2,-p,\sigma_2}(\tau_2)c^*_{k_2'-q_2,-p,\sigma_2'}(\tau_2)
c_{k_2',p,\sigma_2'}(\tau_2)d_{k_2,-p,\sigma_2}(\tau_2) \nonumber \\
&\times&
d^*_{k_3+q_3,-p,\sigma_3}(\tau_3)d^*_{k_3'-q_3,-p,\sigma_3'}(\tau_3)
c_{k_3',p,\sigma_3'}(\tau_3)c_{k_3,p,\sigma_3}(\tau_3)
\rangle_d
+ h.c.\Big\}, \label{Sa} \\
S_b &=& \frac{1}{L^3} \sum_p 
\sum_{\sigma_1,\sigma_1'} \sum_{k_1,k_1',q_1} \int_0^\beta \d \tau_1
2 X_{2\kf, \sigma_1 \sigma_1'}
\sum_{\sigma_2,\sigma_2'} \sum_{k_2,k_2',q_2} \int_0^\beta \d \tau_2
2 X_{2\kf, \sigma_2 \sigma_2'}
\sum_{\sigma_3,\sigma_3'} \sum_{k_3,k_3',q_3} \int_0^\beta \d \tau_1
2 X_{4\kf, \sigma_3 \sigma_3'} \nonumber \\
&\times&
\langle
d^*_{k_1+q_1,p,\sigma_1}(\tau_1)c^*_{k_1'-q_1,-p,\sigma_1'}(\tau_1)
c_{k_1',p,\sigma_1'}(\tau_1)c_{k_1,p,\sigma_1}(\tau_1) \nonumber \\
&\times&
c^*_{k_2+q_2,-p,\sigma_2}(\tau_2)c^*_{k_2'-q_2,-p,\sigma_2'}(\tau_2)
c_{k_2',p,\sigma_2'}(\tau_2)d_{k_2,-p,\sigma_2}(\tau_2) \nonumber \\
&\times&
c^*_{k_3+q_3,-p,\sigma_3}(\tau_3)d^*_{k_3'-q_3,-p,\sigma_3'}(\tau_3)
c_{k_3',p,\sigma_3'}(\tau_3)d_{k_3,p,\sigma_3}(\tau_3)
\rangle_d, \label{Sb} \\
S_c &=& \frac{1}{L^3} \sum_p 
\sum_{\sigma_1,\sigma_1'} \sum_{k_1,k_1',q_1} \int_0^\beta \d \tau_1
2 X_{2\kf, \sigma_1 \sigma_1'}
\sum_{\sigma_2,\sigma_2'} \sum_{k_2,k_2',q_2} \int_0^\beta \d \tau_2
2 X_{2\kf, \sigma_2 \sigma_2'}
\sum_{\sigma_3,\sigma_3'} \sum_{k_3,k_3',q_3} \int_0^\beta \d \tau_1
2 X_{2\kf, \sigma_3 \sigma_3'} \nonumber \\
&\times&
\langle
d^*_{k_1+q_1,p,\sigma_1}(\tau_1)c^*_{k_1'-q_1,-p,\sigma_1'}(\tau_1)
c_{k_1',p,\sigma_1'}(\tau_1)c_{k_1,p,\sigma_1}(\tau_1) \nonumber \\
&\times&
c^*_{k_2+q_2,-p,\sigma_2}(\tau_2)c^*_{k_2'-q_2,-p,\sigma_2'}(\tau_2)
c_{k_2',p,\sigma_2'}(\tau_2)d_{k_2,-p,\sigma_2}(\tau_2) \nonumber \\
&\times&
c^*_{k_3+q_3,-p,\sigma_3}(\tau_3)d^*_{k_3'-q_3,-p,\sigma_3'}(\tau_3)
d_{k_3',p,\sigma_3'}(\tau_3)c_{k_3,p,\sigma_3}(\tau_3)
\rangle_d. \label{Sc}
\end{eqnarray}
In eq.~(\ref{Sa}), the average, $\lan \cdots \ran_d$ is calculated as,
\begin{eqnarray}
&-& \lan d_{k_1,-p,\sigma_1}(\tau_1) d_{k_2,-p,\sigma_2}(\tau_2)
d^*_{k_3+q_3,-p,\sigma_3}(\tau_3) d^*_{k_3'-q_3,-p,\sigma_3'}(\tau_3)
\ran_d \nonumber \\
&=& - G_{-p}(\tau_1 - \tau_3,k_1) G_{-p}(\tau_2 - \tau_3,k_2)
\left( 
\de_{k_3+q_3,k_2} \de_{k_3'-q_3,k_1} 
\de_{\sigma_3,\sigma_2} \de_{\sigma_3',\sigma_1}
-
\de_{k_3+q_3,k_1} \de_{k_3'-q_3,k_2} 
\de_{\sigma_3,\sigma_1} \de_{\sigma_3',\sigma_2}
\right) \nonumber \\
&\simeq& - \frac{1}{8t^2}\delta(\tau_1 - \tau_3) \delta(\tau_2 - \tau_3)
\left( 
\de_{k_3+q_3,k_2} \de_{k_3'-q_3,k_1} 
\de_{\sigma_3,\sigma_2} \de_{\sigma_3',\sigma_1}
-
\de_{k_3+q_3,k_1} \de_{k_3'-q_3,k_2} 
\de_{\sigma_3,\sigma_1} \de_{\sigma_3',\sigma_2}
\right) . 
\label{GF}
\end{eqnarray}
Here the green function, $G_p(\tau,k) = \beta^{-1}
\sum_{\en} {\rm e}^{-\im \en \tau} G_p (\im \en, k)$, 
is replaced by $-1/(2\sqrt{2}t)\delta(\tau)$.
By substituting eq.~(\ref{GF}) into eq.~(\ref{Sa}),
\begin{eqnarray}
S_a &=& \frac{1}{8t^2} \sum_p \sum_{\sigma_1,\sigma_1'} 
\sum_{\sigma_2,\sigma_2'}  
2 X_{2\kf, \sigma_1 \sigma_1'} \times 
2 X_{2\kf, \sigma_2 \sigma_2'} \times
2 X_{4\kf, \sigma_2 \sigma_1} \nonumber \\
&\times& \frac{1}{L^3} \sum_{q_1,q_2,q_3} \int_0^\be \d \tau 
A_{p \sigma_1}(q_1 - q_3) A_{p \sigma_2}(q_2 + q_3)
A_{p \sigma_1'}(-q_1 ) A_{p \sigma_2'}(-q_2),  
\end{eqnarray}
where $A_{p,\sigma}(q) = \sum_k c^*_{k+q,p,\sigma} c_{k,-p,\sigma}$.
In a similar calculation,  one obtains $S_b = S_a$ and
\begin{eqnarray}
S_c &=& \frac{1}{8t^2} \sum_p \sum_{\sigma_1,\sigma_1'} 
\sum_{\sigma_2',\sigma_3}  
2 X_{2\kf, \sigma_1 \sigma_1'} \times 
2 X_{2\kf, \sigma_1 \sigma_2'} \times
2 X_{2\kf, \sigma_3 \sigma_1} \nonumber \\
&\times& \frac{1}{L^3} \sum_{q_1,q_2,q_3} \int_0^\be \d \tau 
A_{p \sigma_1}(q_1 - q_3) A_{p \sigma_1'}(q_2 + q_3)
A_{p \sigma_2'}(-q_1 ) A_{p \sigma_3}(-q_2). 
\end{eqnarray}   
Here we introduce the field operator, 
$\psi_{p, \sigma} = 1/{\sqrt{L}} \sum_k \e^{\im k x} c_{k,p,
\sigma}$ and write the Hamiltonian corresponding to the above action, 
${\cal H}_{1/4}'$, as 
\begin{eqnarray}
{\cal H}_{1/4}'&=& \frac{1}{4t^2} \sum_p \sum_{\sigma_1,\sigma_1'} 
\sum_{\sigma_2,\sigma_2'}  
2 X_{2\kf, \sigma_1 \sigma_1'} \times 
2 X_{2\kf, \sigma_2 \sigma_2'} \times
2 X_{4\kf, \sigma_2 \sigma_1} 
\int \d x 
A_{p \sigma_1} A_{p \sigma_2}
A_{p \sigma_1'} A_{p \sigma_2'} \nonumber \\
&+& \frac{1}{8t^2} \sum_p \sum_{\sigma_1,\sigma_1'} 
\sum_{\sigma_2',\sigma_3}  
2 X_{2\kf, \sigma_1 \sigma_1'} \times 
2 X_{2\kf, \sigma_1 \sigma_2'} \times
2 X_{2\kf, \sigma_3 \sigma_1} 
\int \d x 
A_{p \sigma_1} A_{p \sigma_2'}
A_{p \sigma_1'} A_{p \sigma_3}, 
\label{H14d}
\end{eqnarray}
where $A_{p,\sigma} = \psi^\dagger_{p,\sigma} \psi_{-p,\sigma}$.
The above Hamiltonian can be rewritten as 
${\cal H}_{1/4}' = {\cal H}_{1/4} 
+ {\cal H}^a_{1/4} + {\cal H}^b_{1/4}$, 
where
\begin{eqnarray}
{\cal H}_{1/4}&=& \frac{1}{4t^2} \sum_{p,\sigma} 
\left\{
(2 X_{2\kf, \sigma -\sigma})^2 
(2 X_{4\kf, \sigma \sigma} + 2 X_{4\kf, \sigma -\sigma})
+  
(2 X_{2\kf, \sigma \sigma})^2 2 X_{4\kf, \sigma -\sigma}
\right\} 
\int \d x 
A^2_{p \sigma} A^2_{p -\sigma} \nonumber \\
&+& \frac{3}{8t^2} \sum_{p,\sigma}
2 X_{2\kf, \sigma \sigma} 
(2 X_{2\kf, \sigma -\sigma})^2
\int \d x 
A^2_{p \sigma} A^2_{p -\sigma} \virg 
\label{H14a} \\
{\cal H}^a_{1/4} &=& \frac{1}{4t^2} \sum_{p,\sigma} 
\left\{
2 (2 X_{2\kf, \sigma -\sigma}) (2 X_{2\kf, \sigma \sigma})
( 2 X_{4\kf, \sigma \sigma} + 2 X_{4\kf, \sigma -\sigma} )
\right\} 
\int \d x 
A^3_{p \sigma} A_{p -\sigma} \nonumber \\
&+& \frac{1}{8t^2} \sum_{p,\sigma}
\left\{ 6 X_{2\kf, \sigma - \sigma} 
(2 X_{2\kf, \sigma \sigma})^2
+ (2 X_{2\kf, \sigma - \sigma})^3 
\right\}
\int \d x 
A^3_{p \sigma} A_{p -\sigma} \virg
\label{H14aa} \\
{\cal H}^b_{1/4} &=& \frac{1}{4t^2} \sum_{p,\sigma} 
\left\{
(2 X_{2\kf, \sigma \sigma})^2 (2 X_{4\kf, \sigma \sigma}) 
+ \frac{1}{2} (2 X_{2\kf, \sigma \sigma})^3
\right\} 
\int \d x 
A^4_{p \sigma} \point
\label{H14ab}
\end{eqnarray}   
In the bosonized form, 
eqs.~(\ref{H14a}), (\ref{H14aa}) and (\ref{H14ab}) are proportional 
to $\int  \d x \cos 4 \theta_\rho$,  
$\int  \d x \cos 4 \theta_\rho \cos 2 \theta_\sigma$ 
and $\int  \d x \cos 4 \theta_\rho \cos 4 \theta_\sigma$, respectively, 
and the scaling dimension of the respective term is given by
$2- 8 K_\rho$, $2- 8 K_\rho - 2 K_\sigma$ and $2- 8 K_\rho - 8 K_\sigma$. 
Therefore eq.~(\ref{H14a}) has the largest scaling dimension. 
By substituting $2 X_{2\kf, \sigma \sigma'} = Ua \de_{\sigma, -\sigma'} 
- 2 V_2 a$ and $2 X_{4\kf, \sigma \sigma'} = Ua \de_{\sigma, -\sigma'}
- 2 V_1a + 2 V_2 a $ into eq.~(\ref{H14a}), 
one obtains eqs.~(20a) and (20b).
We note that the coupling constant of the Hamiltonian 
(\ref{H14aa}) is proportional to
$2 X_{2\kf, \sigma - \sigma} = Ua - 2 V_2 a$. 
Equation (\ref{H14ab}) becomes important for  $K_\sigma < 1/4$ 
in the insulating state. 
However, the term, $\cos 2 \theta_\sigma$ is already relevant 
for $K_\sigma < 1$, 
\ie $U < 2 V_2$. 
Then we can safely neglect ${\cal H}^b_{1/4}$.

\section{Spin Susceptibility at $T=0$ for  Hubbard Model with $U>0$}
\begin{figure}[tbl]
\centerline{\epsfxsize=7.0cm\epsfbox{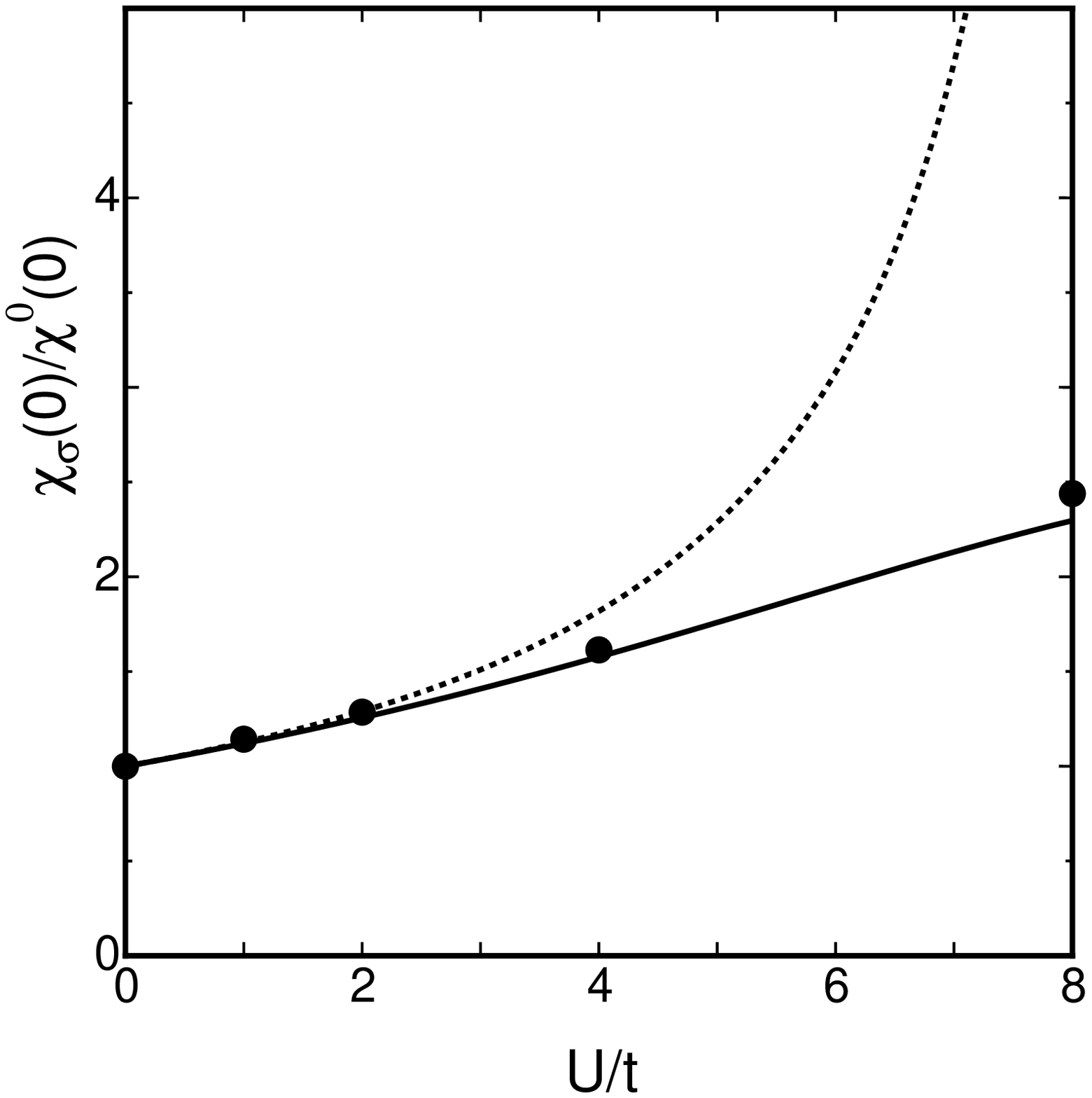}}
\caption{
Spin susceptibility of Hubbard model at $T=0$ 
as a function of $U/t$, where $\chi^0(0)$ is the susceptibility 
in the absence of the interaction. 
The solid (dotted) curve expresses the spin susceptibility
with (without) the virtual process to the upper band shown in Fig.~2. 
The solid circle is derived from the exact solution 
   in ref.~\citen{Shiba}. 
}
\label{fig:chi-comp}
\end{figure}

In case of $V_1 = V_2 = 0$, $U>0$ and $T=0$, 
the spin susceptibility, eqs.~(34) and (35), 
normalized by $\chi^0(0)$
reduces to, 
\begin{eqnarray}
\frac{\chi_\sigma (0)}{\chi^0(0)} &=& \frac{1}
{1 - \displaystyle{\frac{Ua}{2 \pi \vf}}(1 - U D_2)} \point
\end{eqnarray}
We note that the $D_2$-term originates from the virtual process to the
upper band, which is not taken account in ref.~\citen{Nelisse}. 
In Fig.~\ref{fig:chi-comp}, we show $\chi_\sigma (0)/\chi^0(0)$ with and without
the virtual processes 
by solid curve and dotted curve, respectively, which are compared with  
the exact one ( closed circle ).\cite{Shiba}
As seen in Fig.~\ref{fig:chi-comp}, 
the spin susceptibility with the virtual processes is very close  
to the exact solution compared to that without the processes. 
It turns out that the virtual processes 
are necessary for quantitative estimation of  
the spin degree of freedom.



\begin{thebibliography}{99}
\bibitem{CF}
Note that the factor $G_{1/4}(l)$ in the exponential 
of the correlation function 
$\lan \cos 2 \theta_\rho (x) \cos 2 \theta_\rho (0) \ran$
and $\lan \sin 2 \theta_\rho (x) \sin 2 \theta_\rho (0) \ran$
comes from the non-linear term $\cos 4 \theta_\rho$. 
On the hand, the factor does not appear 
in $\lan \cos \theta_\rho (x) \cos \theta_\rho (0) \ran$  
and $\lan \sin \theta_\rho (x) \sin \theta_\rho (0) \ran$.
\end{thebibliography}
\end{document}